\documentclass[structabstract]{aa} 
\usepackage{graphicx}
\usepackage{txfonts}
\usepackage{natbib}
\bibpunct{(}{)}{;}{a}{}{,} % to follow the A&A style

\begin{document}
\title{Gas and dust in the Beta Pictoris Moving Group as seen by the Herschel Space Observatory.\thanks{{\it Herschel} is an ESA space observatory with
    science instruments provided by European-led Principal
    Investigator consortia and with important participation from
    NASA.}}
   \author{P. Riviere-Marichalar\inst{1,2}, D. Barrado\inst{1}, B.Montesinos\inst{1}, G. Duch\^ene\inst{3,4}, H. Bouy\inst{1}, C. Pinte\inst{4}, F. Menard\inst{4,5}, J. Donaldson\inst{6}, C. Eiroa\inst{7}, A. V. Krivov\inst{8}, I. Kamp\inst{2}, I. Mendigut\'ia\inst{9}, W. R. F. Dent\inst{10}, J. Lillo-Box\inst{1}
}

   \institute{Centro de Astrobiolog\'{\i}a (INTA--CSIC) -- Depto. Astrof\'isica, POB 78, ESAC Campus,
	     28691 Villanueva de la Ca\~nada, Spain\\
	     \email{riviere@astro.rug.nl}
   \and Kapteyn Astronomical Institute, P.O. Box 800, 9700 AV Groningen, The Netherlands %2	     
   \and Astronomy Department, University of California, Berkeley CA 94720-3411 USA %3
   \and UJF-Grenoble 1 / CNRS-INSU, Institut de Plan\'{e}tologie et d'Astrophysique (IPAG) UMR 5274, Grenoble, F-38041, France %4	  
   \and Laboratorio Franco-Chileno de Astronomia (UMI 3386: CNRS -- U de Chile / PUC / U Conception),  Santiago, Chile %5
   \and Department of Astronomy, University of Maryland, College Park, MD 230742, USA %6
   \and Dep. de F\'isica Te\'orica, Fac. de Ciencias, UAM Campus Cantoblanco, 28049 Madrid, Spain %7
   \and{Astrophysikalisches Institut und Universit{\"a}tssternwarte, Friedrich-Schiller-Universit{\"a}t Jena, Schillerg{\"a}{\ss}chen 2--3, 07745 Jena, Germany}%8
   \and Department of Physics and Astronomy, Clemson University, Clemson, SC 29634-0978, USA %9
   \and ALMA, Avda Apoquindo 3846, Piso 19, Edificio Alsacia, Las Condes, Santiago, Chile %10
             }
   \authorrunning{Riviere-Marichalar et al.}
   \date{Received 2013 October 23; Accepted 2014 March 27}

 \abstract
%context heading (optional)
{Debris discs are thought to be formed through the collisional grinding of planetesimals, and then can be considered as the outcome of planet formation. Understanding the properties of gas and dust in debris discs can help us comprehend the architecture of extrasolar planetary systems. \textit{Herschel Space Observatory} far-infrared (IR) photometry and spectroscopy have provided a valuable dataset for the study of debris discs gas and dust composition. This paper is part of a series of papers devoted to the study of \textit{Herschel}-PACS observations of young stellar associations.}
{This work aims at studying the properties of discs in the Beta Pictoris Moving Group (BPMG) through far-IR PACS observations of dust and gas. } 
{We obtained \textit{Herschel}-PACS far-IR photometric observations at 70, 100, and 160 $\rm \mu m$ of 19 BPMG members, together with spectroscopic observations for four of them. These observations were centred at 63.18 $\rm \mu m$ and 157 $\rm \mu m$, aiming to detect [OI] and [CII] emission. We incorporated the new far-IR observations in the SED of BPMG members and fitted modified blackbody models to better characterise the dust content.}
{We have detected far-IR excess emission toward nine BPMG members, including the first detection of an IR excess toward HD 29391.The star \object{HD~172555}, shows [OI] emission, while \object{HD~181296} shows [CII] emission, expanding the short list of debris discs with a gas detection. No debris disc in BPMG is detected in both [OI] and [CII]. The discs show dust temperatures in the range 55--264 K, with low dust masses ($\rm  < 6.6 \times 10^{-5}~M_{\oplus}$ to $\rm 0.2 ~M_{\oplus}$) and radii from blackbody models in the range 3 to $\rm \sim $ 82 AU. All the objects with a gas detection are early spectral type stars with a hot dust component.}  
{}

   \maketitle
\section{Introduction} 
Debris discs represent the last stage in the evolution of protoplanetary discs into planetary systems. The regions surrounding the star are replenished with dust through collisional cascades triggered by large bodies inside rings of circumstellar mater that resemble our own asteroid and Kuiper belts. Therefore, debris discs can be considered signposts to the likely existence of planets, or at least of comets \citep[see e. g.][and references therein]{Matthews2014}. Since the discovery of IR excess emission around the main sequence star Vega \citep{Aumann1984}, hundreds of stars with debris discs have been detected, mainly thanks to IRAS, ISO, and \textit{Spitzer Space Telescope} observations. Studies made with \textit{Spitzer Space Telescope} data  revealed a fraction ($\sim 15 \%$) of debris discs around A to K stars \citep{Su2006,Siegler2007,Trilling2007,Hillenbrand2008,Fujiwara2013}.

The properties and evolution of debris discs have been extensively studied over the past decades \citep[][and references therein]{Wyatt2008,Krivov2010}. Most studies focussed on deriving dust disc properties by studying the shape of the spectral energy distribution \citep[SED, see e.g. ][]{Bryden2006,Rebull2008}. Nevertheless, resolved imaging has helped us understand the geometry of debris discs: about 100 of the brightest debris discs have been resolved at various wavelengths, from optical to the sub-mm \citep[see e. g. ][]{Smith1984,Koerner1998,Schneider2006,Smith2009}. They show different structures such as inner gaps, rings, clumps, spiral arms, and asymmetries, which can be signposts to the likely existence of planets \citep[see e. g. ][]{MoroMartin2007}. Although dust has been extensively observed, studies have provided little information about its chemical composition. So far, only a few debris discs have shown solid-state features in their mid-IR spectra, indicating the presence of small and warm grains \citep{Chen2006,Fujiwara2013}. These debris discs are most probably in a transient state, and the small dust grains are produced during periods of intense collisional grinding \citep{Lisse2008,Lisse2009}.

Gas in debris discs has only rarely been detected, although we need to know the total amount of gas present to understand the physical evolution of these systems. Only two debris-disc systems show traces of sub-millimetre CO emission, namely 49 Ceti and HD 21997, implying that the gas abundance in debris discs is low when compared with protoplanetary discs \citep{Dent2005}. Only the $\beta$ Pic debris disc appears to have a full inventory of gaseous species, both atomic and molecular \cite[see e. g.][]{Lagrange1998,Thi2001,Roberge2006}.

The \textit{Herschel Space Observatory} \citep{Pilbratt2010} has produced a valuable dataset for the study of circumstellar environments, allowing both the dust in the continuum and gas emission in the far-IR to be observed. The \textit{Photodetector Array Camera \& Spectrometer}-PACS \citep{Poglitsch2010} can study debris discs with high sensitivity in the wavelength range (55-220 $\mu \rm{m}$) where cold dust emission peaks. PACS spectroscopy allows for detecting a few bright emission lines for molecular species such as CO, $\rm H_{2}O$, and OH and the strong cooling lines of [OI]  and [CII] at 63 and 158 $\mu \rm{m}$, respectively. Some \textit{Herschel} key time programmes, such as DUNES \citep{Eiroa2010} and DEBRIS \citep{Matthews2010},  have focussed on studying debris discs in the far-IR. DUNES \citep[from Dust around NEarby Stars,][]{Eiroa2013}  studied a sample of 124 main sequence stars within 20 pc of the Sun and found an excess due to circumstellar dust in 25 of them (debris-disc rate $\rm \sim 20 \%$), with a mean detection limit in the dust fractional luminosity of $\rm L_{d}/L_{*} \sim 2.0 \times 10^{-6}$. Some of the discs in the DUNES sample show a dust fractional luminosity that is only a few times higher than our Kuiper Belt, but these discs seem to be cooler and larger. Additionally, they discovered a new class of cool debris discs that show excess at 160 $\rm \mu m$ but little or non excess at 100 $\rm \mu m$ \citep[see][]{Krivov2013}.

Young stellar associations \citep{Torres2006} most likely provide the best frame to study the evolution of circumstellar environments, because the ages of their members are well known. In this paper we present \textit{Herschel}-PACS observations of the Beta Pictoris Moving Group (BPMG), a young stellar association with a typical distance of $\rm \sim 30~pc$. The most recent estimation of the age of the system by \cite{Binks2014} ( $\rm 21 \pm 4$ Myr) is in good agreement with the $\rm 20 \pm 10$ Myr age derived by \cite{Barrado1999}. The star that names the group, $\rm \beta$ Pic, is an A6V star that harbours one of the best studied debris discs and that shares its space motion with a group of coeval stars that constitute the BPMG. Nineteen BPMG members were observed with PACS as part of the \textit{Herschel}  Open Time Key Programme GAS in Protoplanetary Systems \citep[GASPS, P.I. W. Dent,][]{Dent2013} that has observed $\rm \sim 250$ stars in different associations with different ages, from Taurus ($\rm 1-3~Myr$) to Tucana Horologium ($\rm \sim$ 30 Myr). Photometry at 70 and/or 100 and 160 $\mu \rm{m}$ was obtained for the 19 stars. Spectroscopic observations were also obtained for four of them. We compared their SEDs with modified blackbody models and derived disc properties, such as inner radii and dust masses based on blackbody modeling results. 

\section{The sample}\label{BPMG:sample}
The BPMG members studied, together with their distances, spectral types, effective temperatures, and luminosities, are listed in Table~\ref{tableStar}. Archival photometry data were collected for each BPMG member in the sample, including Johnson, Stromgren, 2MASS, IRAC, WISE, AKARI, MIPS \citep[from][]{Rebull2008}, LABOCA \citep[from][]{Nilsson2009,Nilsson2010}, and SMA. We also included \textit{Spitzer}-IRS data in the analysis, by collapsing the spectra taken from the archive to  photometric points by using 1 $\mu \rm{m}$ bins, including a 5$\rm \%$ calibration uncertainty  in the errors \citep[see][]{Chen2006}. The different IRS spectral orders were scaled to be consistent with the photospheric flux at short wavelengths when possible. The inclusion of IRS spectra is a central interest, given that the onset of the excess is seen at mid-IR wavelengths. The lack of agreement between IRS data and WISE fluxes made us decide to exclude WISE data from the model analysis in Sec. \ref{dustModels_Sec}.

\begin{table}[!t]
\caption{Stellar parameters for BPMG members observed with \textit{Herschel}-PACS}             
\label{tableStar}      
\centering          
\begin{tabular}{l l l l l}     % 6 columns 
\hline\hline       
Name & d & Sp. type & T$_{eff}$ &  L$_*$  \\ 
--	& (pc) & --  & (K) & L$_{sun}$  \\ 
\hline                    
\object{AT~Mic} &  10.7 & M4V & 3100 &  0.066  \\
\object{CD64-1208} & 29.2 & K5V  & 4200 & 0.25 \\
\object{GJ~3305} & 29.8 & M1V  & 3600 & 0.166 \\
\object{HD~203} & 39.1 & F3V & 6600 & 3.7  \\
\object{HD~29391} & 29.8 & F0V & 7400 & 5.43  \\
\object{HD~35850} & 26.8 & F7V & 6000 & 1.26  \\
\object{HD~45081} & 38.5 & K4V &  4200  & 0.329  \\
\object{HD~139084} & 39.8 & K0V & 5000 & 1.23 \\
\object{HD~146624} & 43.1 & A0V & 9750 & 22 \\
\object{HD~164249} & 46.9 & F6V &6600 & 2.7 \\
\object{HD~172555} & 29.2 & A6IV & 7800 & 7.8 \\
\object{HD~174429} & 49.7 & G9IV & 5200 & 0.85 \\
\object{HD~181296} & 47.7 & A0V & 10000 & 23.1  \\
\object{HD~181327} & 50.6 & F6V & 6600 & 3.2  \\
\object{HD~199143} & 47.7 & F7V & 6000 & 1.93  \\
\object{HIP~10679} & 34.0 & G2V & 5800 & 0.87  \\
\object{HIP~10680} & 39.4 & F5V & 6200 & 2.4 \\
\object{HIP~11437} & 42.3 & K6V & 4400 & 0.242  \\
\object{HIP~12545} & 40.5 & K6V & 4000 & 0.238  \\
\hline                  
\end{tabular}
\tablefoot{Spectral types are taken from \cite{Zuckerman2001,Zuckerman2004,Torres2006}. Stellar distances are taken from \cite{Zuckerman2004}}
\end{table}

\begin{figure}[!b]
\begin{center}
%   \centering
     \includegraphics[scale=0.5]{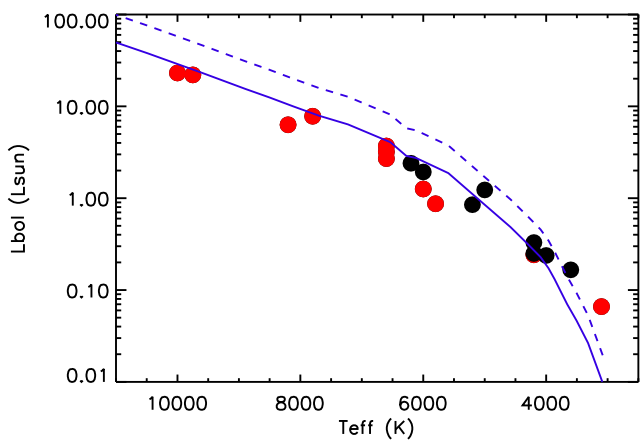}
   \caption{Hertzsprung-Russell diagram for BPMG members in the sample. The blue solid line represents the 20 Myr isochrone, while the blue dashed one represents a 20 Myr isochrone with 2 times larger luminosity to include unresolved binary systems. The isochrones come from the models by \cite{Siess2000}. Red filled circles are objects detected by PACS while black filled circles are objects not detected in PACS photometry.}
   \label{BPMG_HRD}
\end{center}
\end{figure}

The first step before modelling the whole SED is to derive the stellar parameters. For each star in the sample, we selected photometric points with a pure photospheric origin (typically in the 0.3--3.0 $\rm \mu m$ range) and used them to compare the non-excess SED with theoretical models using the Virtual Observatory SED Analyzer\footnote{http://svo2.cab.inta-csic.es/theory/vosa/} \citep[VOSA,][]{Bayo2008}. VOSA compares the photometric data with a grid of models to match the photosphere that produces the best fit, based on a $\rm \chi^{2}$ minimisation, and, if desired, on Bayesian analysis. The values for $\rm T_{eff}$ and $\rm L_{*}$ from the best fits are summarised in Table~\ref{tableStar}. In Fig. \ref{BPMG_HRD} we plot the Hertzsprung-Russell diagram  for the stars in the sample.

\section{Observations and data reduction}
Nineteen BPMG members were observed in photometric mode and a subset of four BPMG members were also observed in spectral line (63/189 $\rm \mu m$ ) and range mode (78/157 $\rm \mu m$; see Table \ref{obsLog} for an overview of the observations containing the PACS OBSIDs).

\begin{table*}[!t]
\caption{Obs IDs for BPMG member observed with \textit{Herschel}-PACS}             
\label{obsLog}      
\centering          
\begin{tabular}{l l l l l}     % 6 columns 
\hline\hline       
Name & Photometry OBSID & Wavelength$\rm ^{a}$ & Spectroscopy OBSID & Spectral Ranges  \\ 
-- & -- & ($\rm \mu m$) & -- & ($\rm \mu m$) \\ 
\hline                    
\object{AT~Mic} &  1342196104, 1342209488, 1342209489 & 70, 70, 100 & -- &--\\
\object{CD64-1208} & 1342209059, 1342209060, 1342209061 & 70,100, 70, 100 & -- & --\\
\object{GJ~3305} &  1342224850, 1342224851, 1342224852, 1342224853 & 70, 70, 100, 100 & --& --\\
\object{HD~203} &  1342188366, 1342221118, 1342221119 & 70, 100, 100 & -- & --\\
\object{HD~29391} & 1342190967, 1342216153, 1342216154 & 70, 100, 100 & -- & -- \\ 
\object{HD~35850} & 1342217746, 1342217747 & 70, 100 & -- & --\\
\object{HD~45081} &  1342188506, 1342212832, 1342212833 & 100, 70, 70 & -- & --\\
\object{HD~139084} & 1342216483, 1342216484, 1342216485, 1342216486 & 70, 70, 100, 100 & -- & --\\
\object{HD~146624} & 1342215617, 1342215618, 1342215619, 1342215620 & 70, 70, 100, 100 & --  & --\\
\object{HD~164249} & 1342183657, 1342215574, 1342215575 & 70, 100, 100 & 1342215648, 1342239388 & 63, 157 \\
\object{HD~172555} &1342209059, 1342209060, 1342209061 & 70,100, 70, 100 & 1342215649, 1342228416, 1342228417 & 63, 157, 63 \\
\object{HD~174429} & 1342215576, 1342215577, 1342215578, 1342215579 & 70, 70, 100, 100 & -- & -- \\
\object{HD~181296} & 1342290955, 1342209056 & 100, 100, & 1342209730, 1342239756 & 63, 157 \\
\object{HD~181327} & 1342183658, 1342209057, 1342209058 & 70, 100, 100 & 1342186311, 1342186810 & 63, 157 \\
\object{HD~199143} &  1342193550, 1342208861, 1342208862 & 70, 70, 70 & -- & --  \\
\object{HIP~10679} & 1342189193, 1342223862, 1342223863 & 70, 100, 100 &  -- & --  \\
\object{HIP~10680} & 1342189193, 1342223862, 1342223863 & 70, 100, 100 &  -- & --  \\
\object{HIP~11437} & 1342189210, 1342223864, 1342223865 & 70, 100, 100 & -- & -- \\
\object{HIP~12545} & 1342189150, 1342223574, 1342223575 & 70, 100, 100 & -- & --\\
\hline                  
\end{tabular}
\tablefoot{(a): PACS photometer simultaneously observes in either the70 or 100 $\rm \mu m$ bands plus the 160 $\rm \mu m$ band (see Sec. \ref{Sec:PhotRed}).}
\end{table*}

\subsection{Photometric observations and data reduction}\label{Sec:PhotRed}
The PACS photometer simultaneously observes in either the 70 or 100 $\mu \rm{m}$ band, together with the 160 $\mu \rm{m}$ band, so when the source have been observed in both 70 and 100 $\mu {\rm m}$, we have two images to combine in the 160 $\mu \rm{m}$ band. We observed 16 objects at 70 $\mu {\rm m}$ and another 16 at 100 $\mu {\rm m}$, making a total of 19 observations in the 160 $\mu {\rm m}$ band (see Table~\ref{obsLog}). The exposure times range from 133 to 1122 s, based on the expected flux from the star. Each scan map was made with medium speed ($\rm 20~$\arcsec$s^{-1}$), with scan legs of 3 arcmin and a separation of 4-5 $\arcsec$ between legs.

We reduced the photometric observations using HIPE 8, with the most recent version of the calibration files, in the same fashion as our TW Association data \citep{Riviere2013}. For bright IR-excess targets (i. e., those with flux greater than $\rm \sim$ 100 mJy), we used a version of the pipeline tuned for bright objects, while for faint objects and non-detected objects we used a different version optimised for noise-dominated maps. Both pipelines shared the following reduction steps: bad and saturated pixel flagging and removal, flat field correction, deglitching, high pass filtering, and map projection. We refer the reader to \cite{Riviere2013} for a detailed description of the differences between both pipelines. Photometric maps were projected into the final image with pixel scale 2 $\arcsec$/pixel in the 70 and 100 $\rm \mu m $ bands and with pixel scale 3 $\arcsec$/pixel in the 160 $\mu \rm{m}$ band. We also produced final maps with the native pixel scale of the detector that were used to perform the error calculation (3.2 $\arcsec$/pixel for the 70/100 $\mu \rm{m}$ bands and 6.4 for the 160 $\mu \rm{m}$ band). When several images at the same wavelength are available for a single target, we combined all of them to improve the signal-to-noise ratio (S/N), averaging for each pixel and using the average sigma clipping algorithm to exclude bad pixels. 

Aperture photometry was measured using an aperture of 6$\arcsec$ for the 70 and 100 $\mu \rm{m}$ bands and 12$\arcsec$ for the 160 $\mu \rm{m}$ band. The annulus for sky subtraction was placed at 25-35$\rm \arcsec$ from the star. We then applied an aperture correction for each band\footnote{http://herschel.esac.esa.int/twiki/pub/Public/PacsCalibrationWeb/}. Final fluxes are listed in Table~\ref{HSOphot}. Noise errors consist of the standard deviation of the photometry obtained at several sky positions surrounding the target. PACS calibration uncertainties are 2.64, 2.75, and 4.15 $\%$ for the 70, 100, and160 $\mu \rm{m}$ bands, respectively. Noise errors and calibration errors were added quadratically. 

For non-detected sources, we derive upper limits as follows: we compute aperture photometry in the sky background in several pointings that surround the nominal position of the star on the detector. Then we compute the standard deviation, and we use it as the sky background value. The upper limits included in Table~\ref{HSOphot} are 3$\rm \sigma$.

\subsection{Spectroscopic observations and data reduction}
Four BPMG members (HD 164249, HD172555, HD 181296, and HD 181327) were observed with PACS in LineScan spectroscopic mode, targeting [OI] emission at 63.18 $\mu \rm{m}$. \cite{Riviere2012} reported the detection of o-$\rm H_{2}O$ emission at 63.32 $\mu \rm{m}$ in the PACS spectra of eight T Tauri stars in Taurus, therefore we also searched for line emission at this wavelength. All the targets were also observed in RangeScan mode, aiming to detect [CII] emission at 157.74 $\rm \mu m$. PACS spectra were reduced using HIPE 9 with the latest version of the pipeline and the proper calibration files. Spectra were extracted from the central spaxel and aperture-corrected to account for flux spread in the surrounding spaxels. The line spectra from PACS typically show lower S/N near the spectrum edges. To account for that effect, we exclude from the spectra any wavelength shorter than 63.0 $\mu \rm{m}$ or longer than 63.4 $\mu \rm{m}$ in the 63 $\mu \rm{m}$ range. In the RangeScan observations we exclude wavelengths shorter than 157 $\rm \mu m$ and longer than 159 $\rm \mu m$.

Line fluxes were computed by applying a Gaussian fit to the line profile and calculating the integrated flux from that fit. Upper limits were computed by integrating a Gaussian with a width equal to the instrumental FWHM at the central wavelength, and maximum equal to three times the standard deviation of the continuum (upper limits are 3$\rm \sigma$). HD 172555 was observed at two different epochs in LineScan mode, the emission being detected in both observations. When combining both epochs, small shifts in the line centres translate into an increase in noise. To avoid this effect, we computed line fluxes for the averaged spectrum after re-centring both spectra by artificially shifting them in a way that the line centre from the fit is exactly at the rest frame wavelength of the observed line. Line fluxes are shown in Table~\ref{HSOspec}.

\begin{table*}[!ht]
\caption{\textit{Herschel}/PACS photometry and estimated fluxes from the naked photosphere}             
\label{HSOphot}      
\centering          
\begin{tabular}{c c c c c c c}     % 6 columns 
\hline\hline       
Name & $\rm F_{\nu , ~obs}$(70 $\mu \rm{m}$)  & $\rm F_{\nu , ~phot}$(70 $\mu \rm{m}$)  & $\rm F_{\nu , ~obs}$(100 $\mu \rm{m}$)  & $\rm F_{\nu , ~phot}$(100 $\mu \rm{m}$) &  $\rm F_{\nu , ~obs}$(160 $\mu \rm{m}$)  & $\rm F_{\nu , ~phot}$(160 $\mu \rm{m}$)  \\ 
--	& (mJy)  & (mJy)  & (mJy) & (mJy)  & (mJy)  & (mJy) \\ 
\hline                    
\object{AT~Mic} &  13 $\pm$ 2 & 15$\pm$3 & -- & 7.3$\pm$0.8 & $<$ 14 & 2.8$\pm$0.3 \\
\object{CD64-1208} & $<$ 8 & 3.2$\pm$0.4 & $<$ 9 & 1.6$\pm$0.2 & $<$ 8 & 0.65$\pm$0.08 \\
\object{GJ~3305} & $<$ 4 & 3.1$\pm 0.5$ & $<$ 4 & 1.5$\pm$0.2 & $<$ 8 & 0.6 \\
\object{HD~203} & 68 $\pm$ 3 & 6.5$\pm$0.4 & 26 $\pm$ 2 & 3.2$\pm$0.2 & $<$ 14 & 1.24$\pm$0.08 \\
\object{HD~139084} & $<$ 7 & 4.2$\pm$ 0.5& $<$ 10 & 2.0$\pm$0.2 & $<$ 29 & 0.81$\pm$0.09 \\
\object{HD~146624} & 13 $\pm$ 2 & 9.1$\pm$0.2 & $<$ 7 & 4.47$\pm$0.09 & $<$ 13 & 1.75$\pm$0.03 \\
\object{HD~164249} & -- & 3.4$\pm$0.2 &  493 $\pm$ 20 & 1.66$\pm$0.07 & 249 $\pm$ 10 & 0.65$\pm$0.03 \\
\object{HD~172555} & 191 $\pm$ 7 & 16.5$\pm$0.5 & 79 $\pm$ 4 & 8.0$\pm$0.3 & 31 $\pm$ 2 & 3.2$\pm$0.1 \\
\object{HD~174429} & $<$ 7 & 2.6$\pm$0.4 & $<$ 7 & 1.3$\pm$0.2 & $<$ 13 & 0.58$\pm$0.07\\  
\object{HD~181296} & -- & 7.8$\pm$0.2 & 250 $\pm$ 8 & 3.81$\pm$0.07 & 111 $\pm$ 6 & 1.49$\pm$0.03 \\
\object{HD~181327} & 1453 $\pm$ 6 & 3.4$\pm$0.2 & 1400 $\pm$ 40 & 1.62$\pm$0.9 & 850 $\pm$ 40 & 0.63$\pm$0.04\\
\object{HD~199143} & $<$ 5 & 3.6$\pm$0.4 & -- & 1.8$\pm$0.2 & $<$ 11 & 0.71$\pm$0.07 \\ 
\object{HD~29391} & 22 $\pm$ 2 & 12.6$\pm$0.8 & 17 $\pm$ 3 & 6.2$\pm$0.4 & $<$ 16 & 2.4$\pm$0.2 \\ 
\object{HD~35850} & -- & 8.6$\pm$0.6 & 42 $\pm$ 2 & 4.2$\pm$ 0.3 & 18 $\pm$ 4 & 1.7$\pm$0.1 \\
\object{HD~45081} & $<$ 13 & 1.8$\pm$0.2 & $<$ 10 & 0.9$\pm$0.1 & $<$16 & 0.35$\pm$0.05 \\
\object{HIP~10679} & 53 $\pm$ 3 & 2.7$\pm$0.3 & 46 $\pm$ 2 & 1.3$\pm$0.1 & 39 $\pm$ 5& 0.51$\pm$0.05\\ 
\object{HIP~10680} & $<$ 13 & 4.4$\pm$0.3 & $<$ 8 & 2.2$\pm$0.02 & $<$ 22 & 0.85$\pm$0.06 \\ 
\object{HIP~11437} & 70 $\pm$ 4 & 1.34$\pm$0.03 & 69 $\pm$ 3  & 0.65$\pm$0.02 & 50 $\pm$ 4 & 0.255$\pm$0.006 \\ 
\object{HIP~12545} &  $<$ 7 & 1.5$\pm0.2$ & -- & 0.7$\pm$0.1 & $<$ 19 & 0.29$\pm$0.05 \\ 
\hline                  
\end{tabular}
\end{table*}

\section{Results}
\subsection{Herschel/PACS photometry} 

Photometry results are shown in Table~\ref{HSOphot}. Following \cite{Bryden2006}, we used the parameters $\rm \chi_{70}$, $\rm \chi_{100}$, and $\rm \chi_{160}$ to identify excess sources, defined as
\begin{equation}
\rm \chi_{band} = {F_{obs,band} - F_{*,band} \over \sigma_{band}}
\end{equation}
where $\rm F_{obs,band}$ is the observed flux at any of the 70, 100, or 160 $\mu \rm{m}$ bands, $\rm F_{*,band}$ is the expected photospheric flux at the corresponding wavelength, and $\rm \sigma_{band}$ is the corresponding error, which is the quadratic sum of the photometric error and the photosphere model error. To compute the error from the models, we produced models with temperatures that are the best fit temperature $\rm \pm ~200 ~K$ for each star and derived the errors as the mean difference between the best fit model and those models. We consider that excess sources at each band are those with $\rm \chi_{band}> 3$. 

The agreement between MIPS photometry \citep[from][]{Rebull2008}  and PACS photometry is generally good, with a mean difference smaller than $\rm 10\%$. We detected eight objects at 70 $\mu \rm{m}$ out of 16 observed. Two of them, namely \object{AT~Mic} and \object{HD~146624}, showed fluxes in agreement with photospheric emission ($\chi_{70} \rm{\simeq-0.7}$ and 2.1, respectively). \cite{Plavchan2009} report the MIPS flux at 24 $\rm \mu m$ toward AT Mic to be in excess over the photosphere; however, they predict a photospheric flux at 24 $\rm \mu m$ of 114 mJy, while we predict a flux of 126 mJy ($\rm \chi_{24} \sim 2.7$). Nevertheless, the authors did not include model errors when computing  $\rm \chi_{24}$: if we include the model error at 24 $\rm \mu m$, we get $\rm \chi_{24} \sim 0.4$. We conclude that the 24 $\rm \mu m$ flux is compatible with pure photospheric emission. At 70 $\rm \mu m$, we measure a flux of $\rm (13 \pm 2)$ mJy, compared to the value of $\rm 22 \pm 5$ mJy by \cite{Plavchan2009}: we speculate that the larger MIPS flux is due to the larger MIPS beam size that results in a higher background pollution. \object{HD~29391} showed a flux of $\rm 22 \pm 2~mJy$ at 70 $\mu \rm{m}$ and a flux of $\rm 17 \pm 3~mJy$ at 100 $\mu \rm{m}$, while the expected photospheric fluxes are $\rm \sim 13~mJy$ and $\rm \sim 6.2~mJy$, respectively. The difference between the expected and the observed flux is more than 3$\rm \sigma$ in both bands ($ \chi_{70} \rm{\simeq 4.6}$, $ \chi_{100} \rm{\simeq 3.6}$), and therefore \object{HD~29391} has an infrared excess.

\begin{figure}[!hd]
\begin{center}
%   \centering
      \includegraphics[scale=0.45]{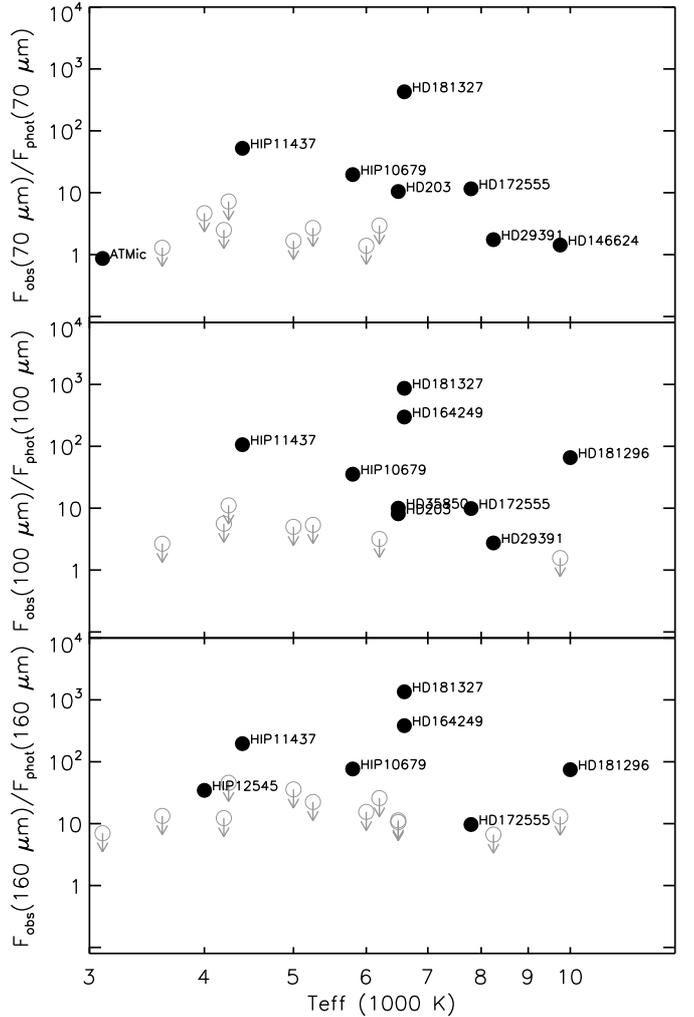}
   \caption{Excess for the 70/100/160 $\mu \rm{m}$ band versus $\rm T_{eff}$. Filled dots represent true detections, while empty circles with arrows represent 3$\sigma$ upper limits}
   \label{ExcessVSTeff}
\end{center}
\end{figure}

\begin{figure*}[!ht]
\begin{center}
   \centering
     \includegraphics[trim=0mm 22mm 0mm 0mm,clip,scale=0.5]{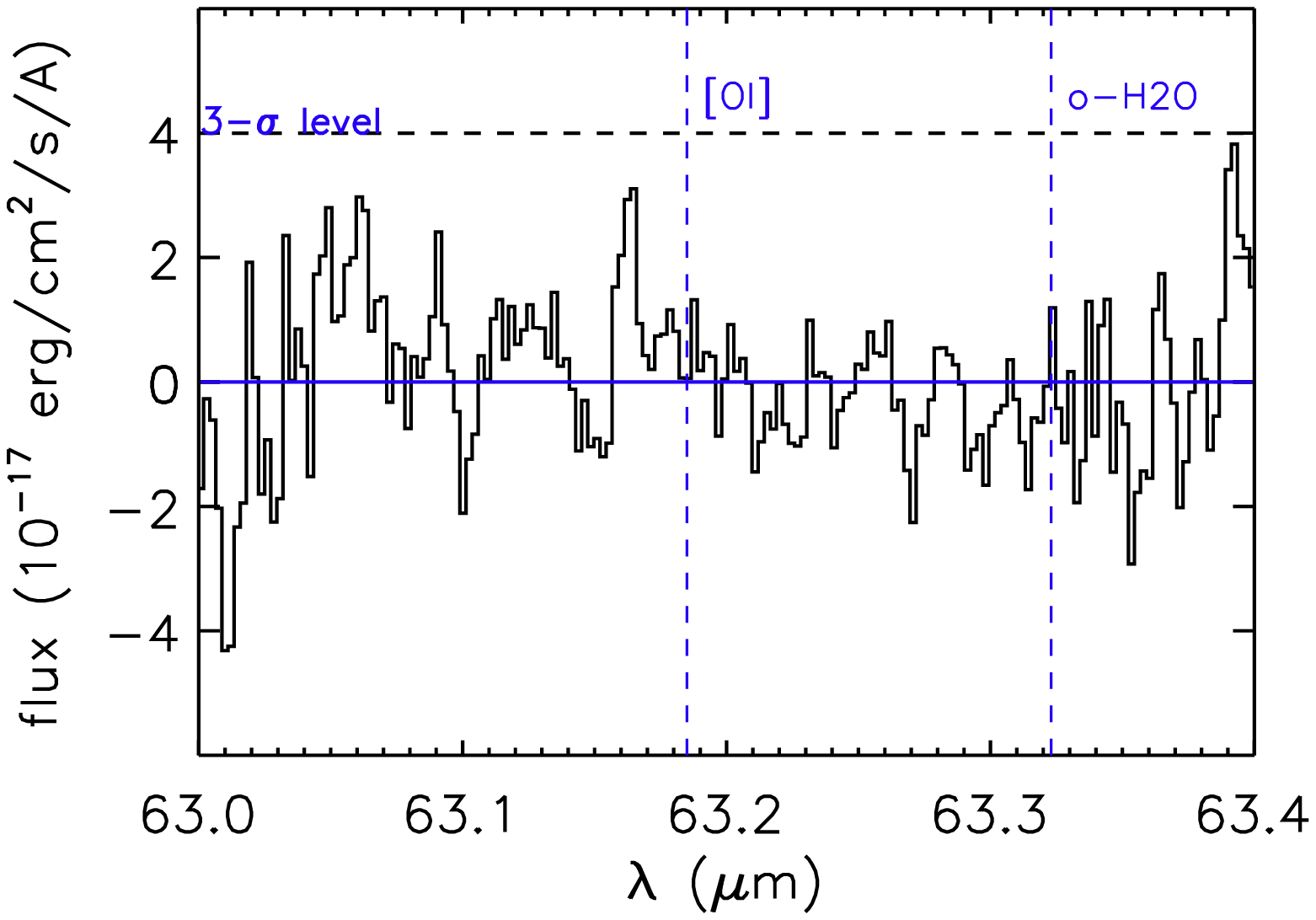}\includegraphics[trim=28mm 22mm 0mm 0mm,clip,scale=0.5]{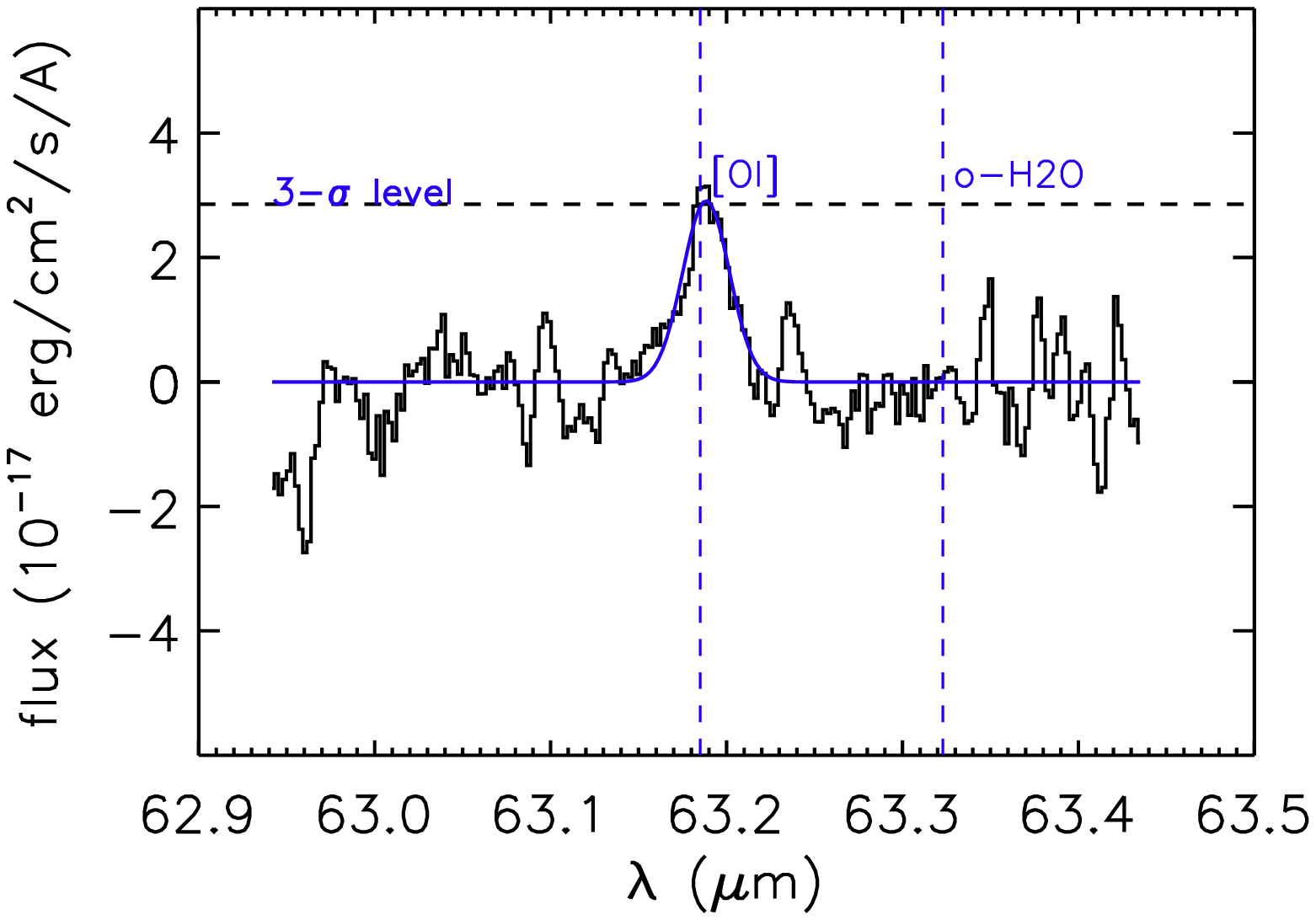}\\
     \includegraphics[trim=0mm 0mm 0mm 0mm,clip,scale=0.5]{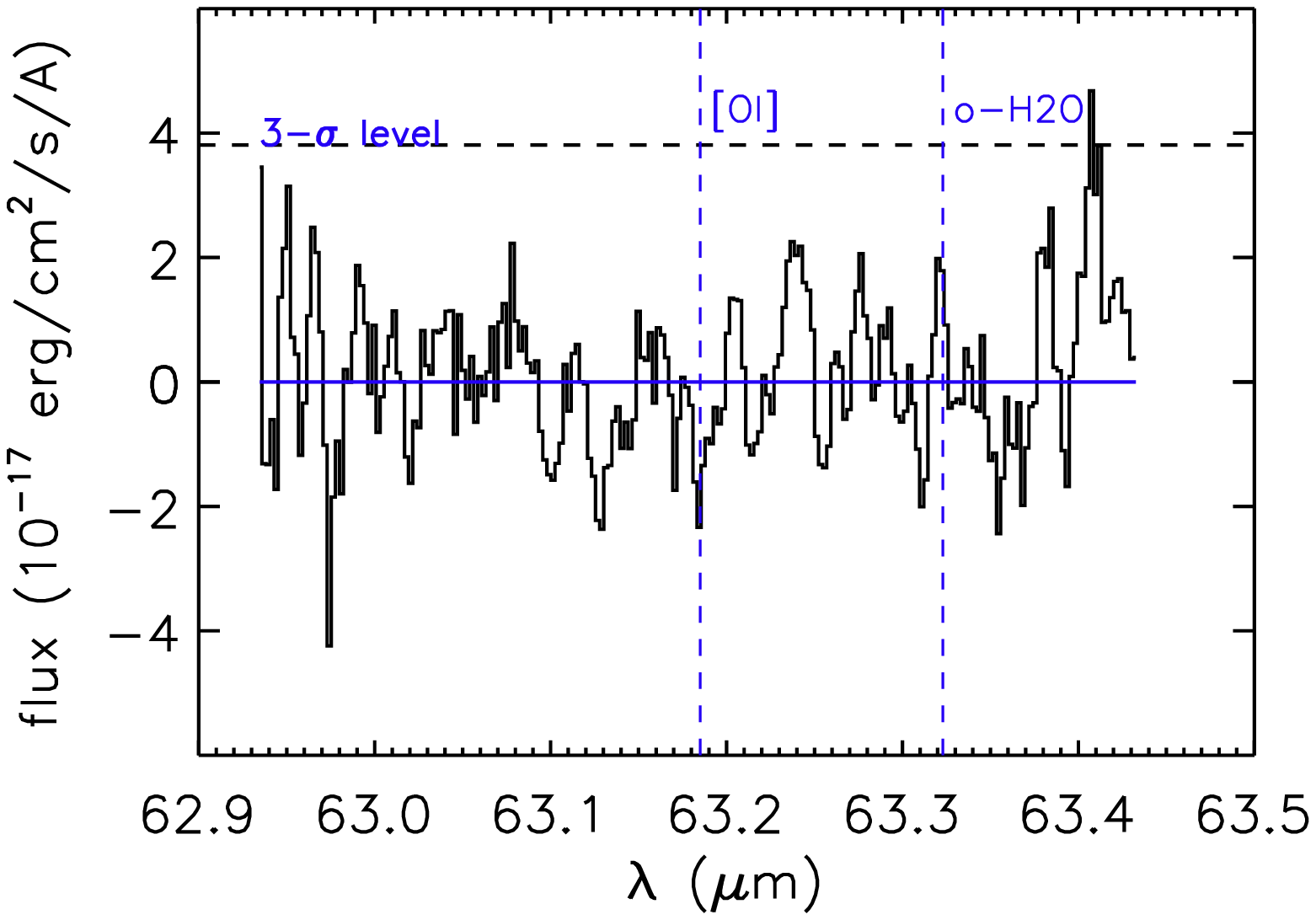}\includegraphics[trim=28mm 0mm 0mm 0mm,clip,scale=0.5]{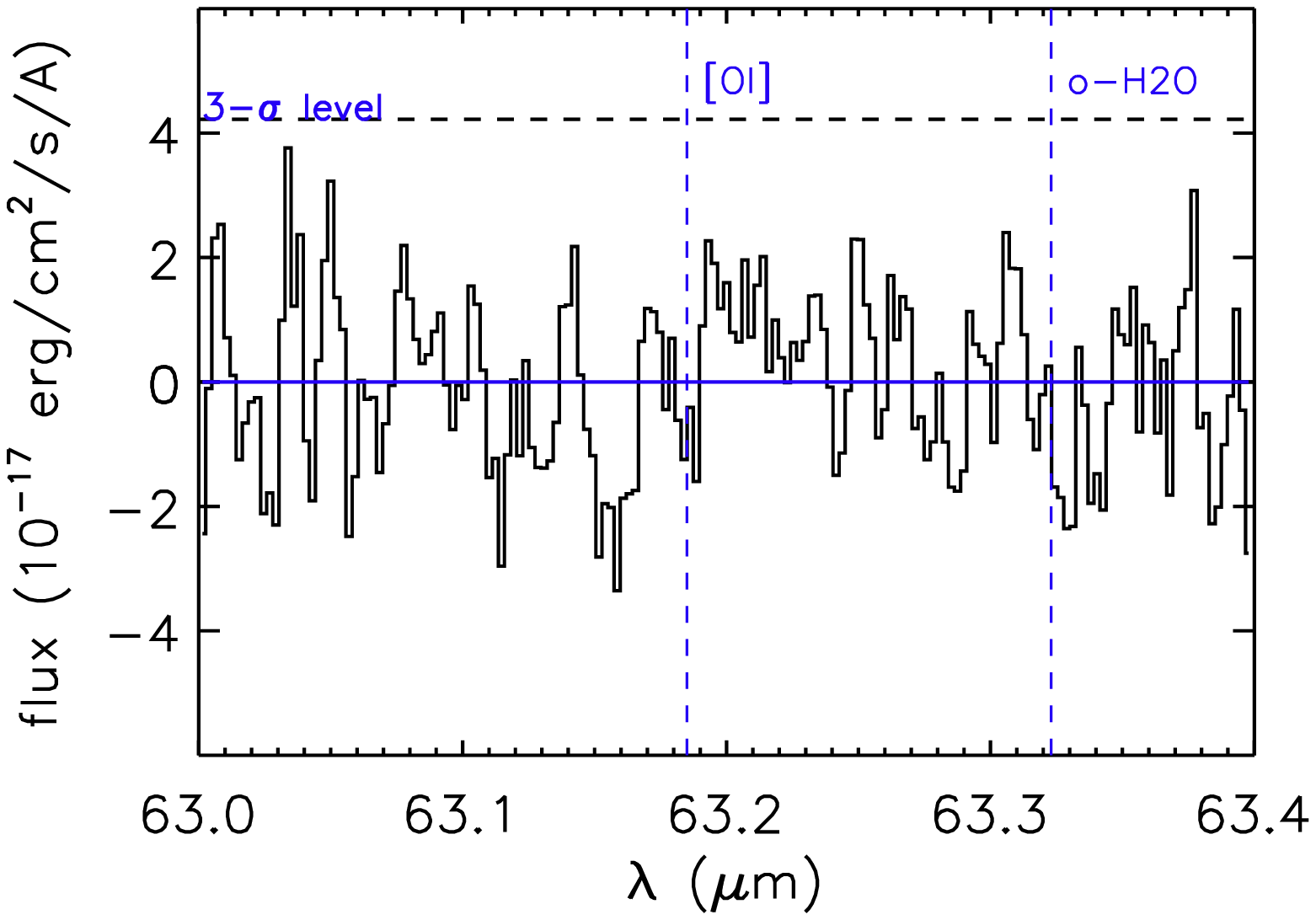}
   \caption{Continuum-subtracted spectra of BPMG members at 63 $\mu \rm{m}$. The vertical blue dashed lines represent the position of the [OI] and o-$\rm H_{2}O$. From left to right and top to bottom, targets are: HD 164249, HD 172555 (averaged and re-centred with respect to the rest frame wavelength), HD 181296, and HD 181327. We show 3$\rm \sigma$ limits as horizontal black dashed lines. For HD 172555 we show in blue a Gaussian fit to the data, while for the other sources the blue horizontal line depicts the position of the continuum.}
   \label{LineSpec63}
\end{center}
\end{figure*}

\begin{figure*}[!ht]
\begin{center}
   \centering
     \includegraphics[trim=0mm 22mm 0mm 0mm,clip,scale=0.5]{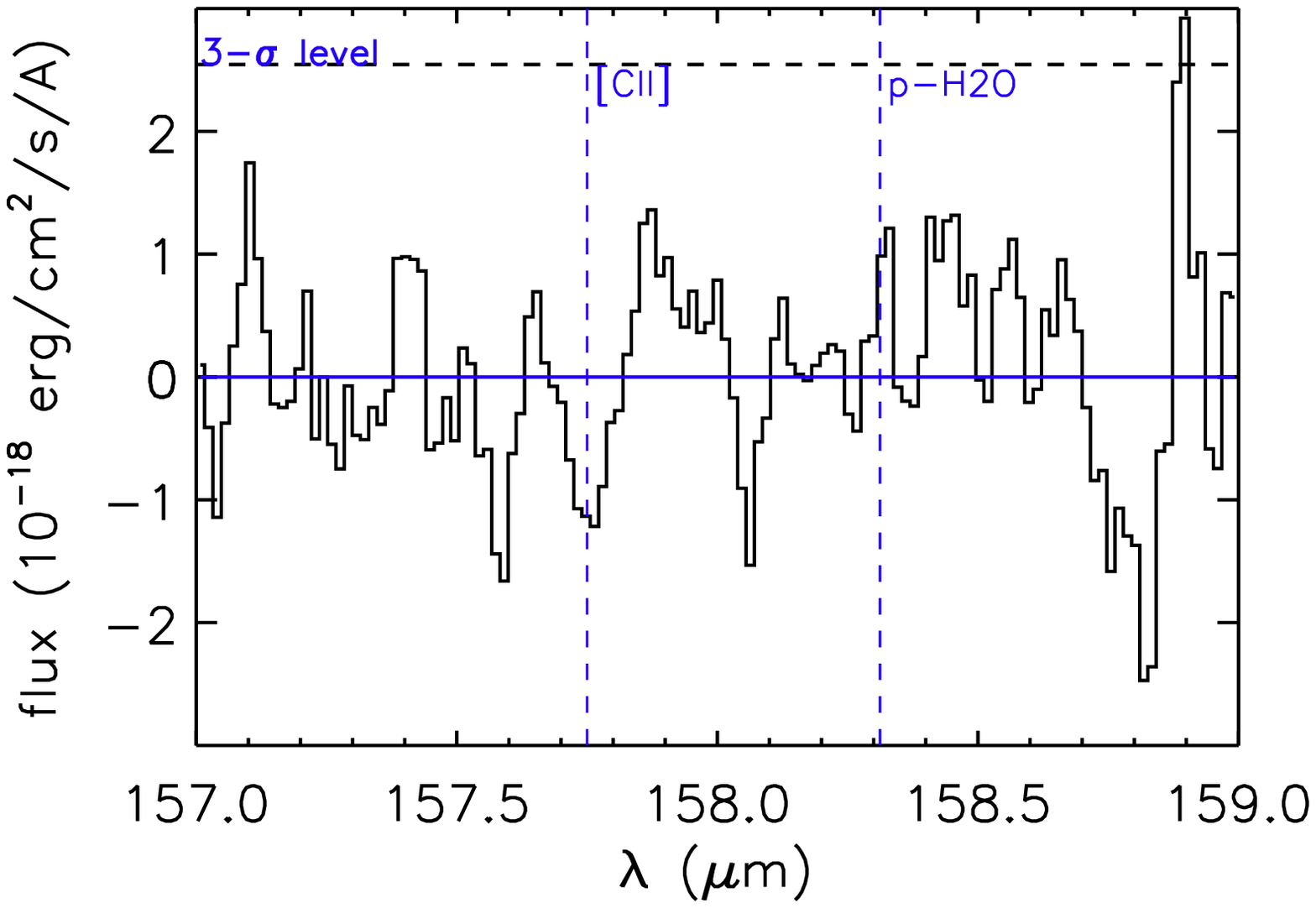}\includegraphics[trim=28mm 22mm 0mm 0mm,clip,scale=0.5]{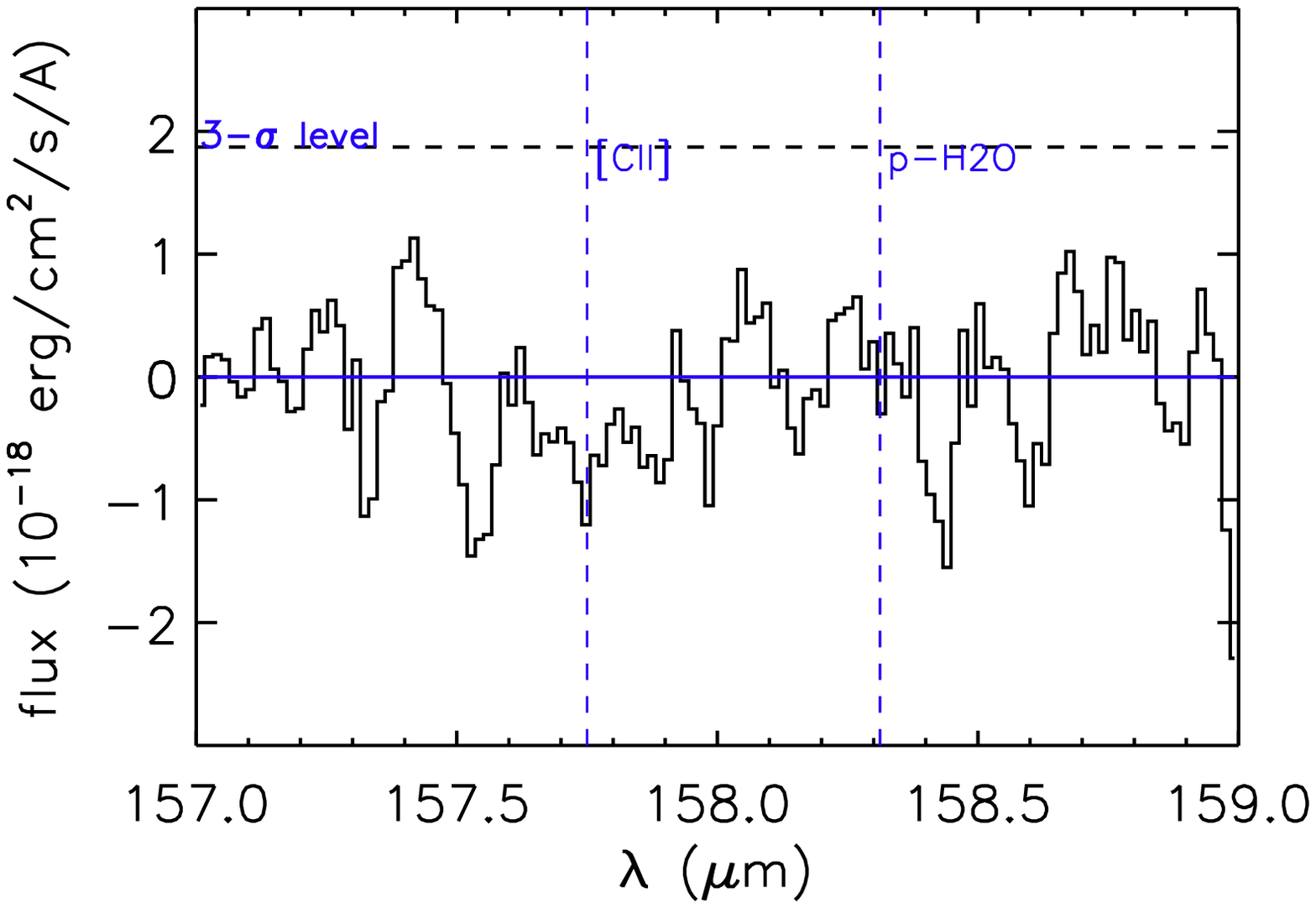}\\
     \includegraphics[trim=0mm 0mm 0mm 0mm,clip,scale=0.5]{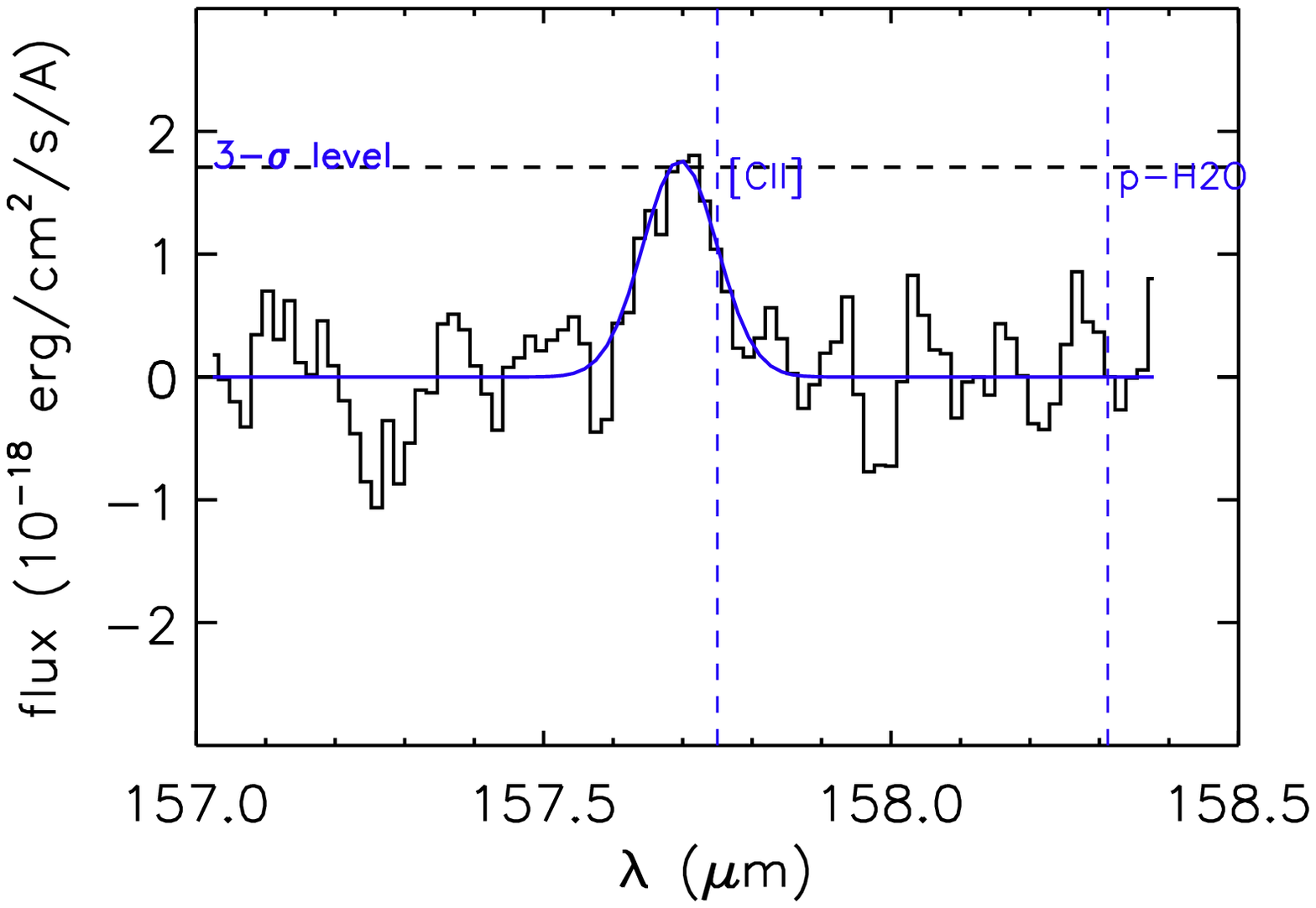}\includegraphics[trim=28mm 0mm 0mm 0mm,clip,scale=0.5]{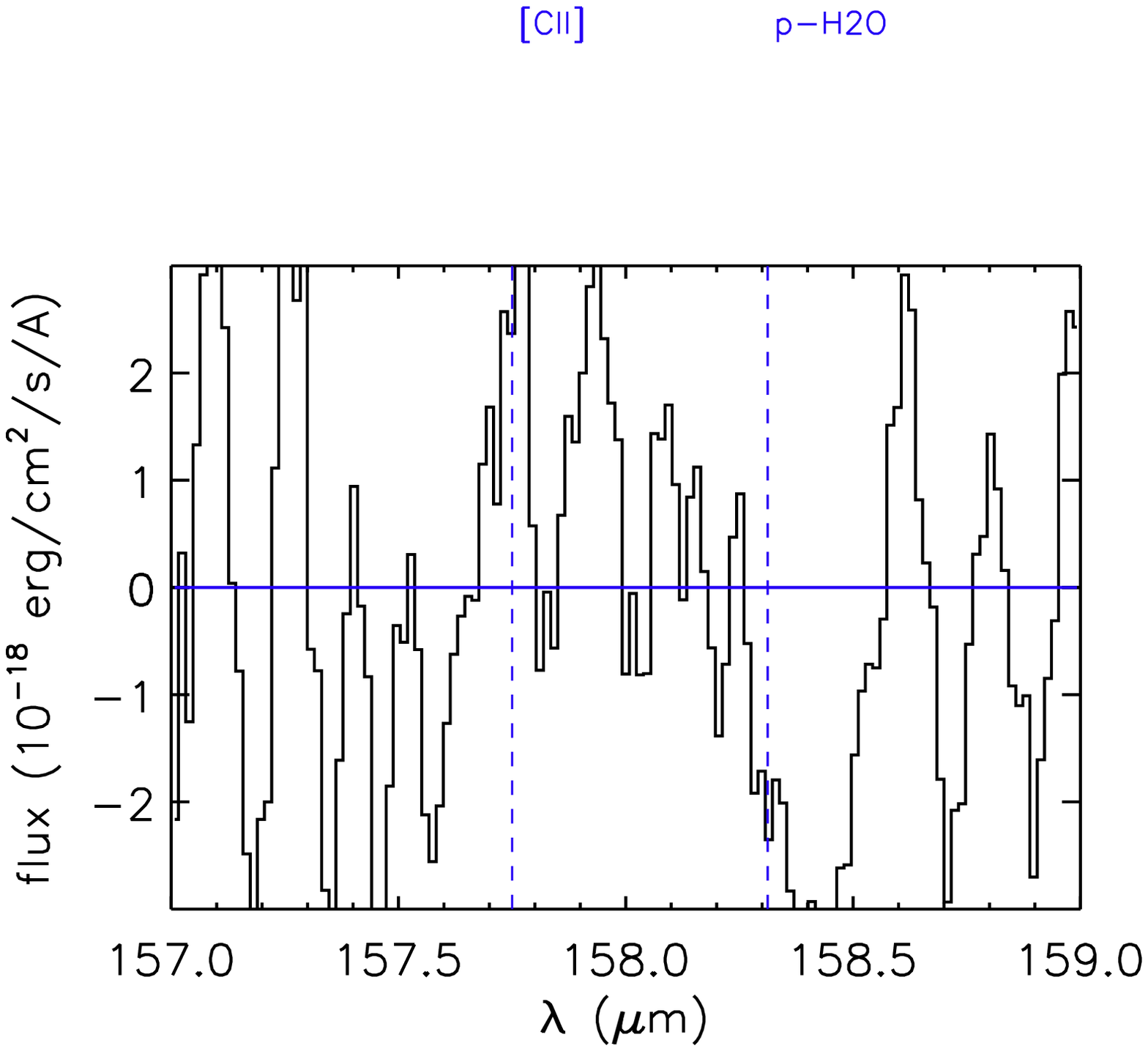}
   \caption{Continuum-subtracted spectra of BPMG members at 157 $\mu \rm{m}$. The vertical blue dashed lines represent the position of the [OI] and o-$\rm H_{2}O$. From left to right and top to bottom, targets are HD 164249, HD 172555 (averaged and re-centred with respect to the rest frame wavelength), HD 181296, and HD 181327. We show 3$\rm \sigma$ limits as horizontal black dashed lines. For HD 181296 we show in blue a Gaussian fit to the data, while for the other sources the blue horizontal line depicts the position of the continuum. The noise in HD 181327 spectrum is more than two times greater than in the other spectra.}
   \label{LineSpec157}
\end{center}
\end{figure*}

In Fig. \ref{ExcessVSTeff} we show the ratio of the observed fluxes to the photospheric fluxes at 70/100/160 $\mu \rm{m}$ versus the effective temperature of the star. Objects not detected at 70 $\mu \rm{m}$ show upper limits that are one to ten times larger than the expected photospheric value. Upper limits at 100 $\mu \rm{m}$ also range from one to ten times the photospheric value. Upper limits at 160 $\mu \rm{m}$ are one to 100 times greater than the photospheric value. We can't exclude the presence of very cold, faint discs for some of the non-detected sources. We find no correlation between the strength of the IR excess at any of the PACS bands and the temperature of the central object, as other authors have reported \citep[see][]{Trilling2008,Eiroa2013}. 

To study whether we have spatially resolved the sources, we performed azimuthally averaged radial profiles in all the three PACS bands for every detected object and compared the results with the azimuthally averaged radial profile of a model PSF source. Only \object{HD~181327} seems to be more extended than the reference star $\alpha$ Boo in the 100 $\mu \rm{m}$ band, with a FWHM of 7.58$\rm \arcsec$, compared to 6.94$\rm \arcsec$ for $\alpha$ Boo, a result already reported by \cite{Lebreton2012}.

\subsection{Herschel/PACS spectroscopy}\label{PACSspecResults}
Among the four BPMG members observed with PACS in LineScan mode, we detected the continuum level at 63 $\mu \rm{m}$ for three of them, \object{HD~164249} being the only exception. Continuum subtracted spectra at 63 $\mu \rm{m}$ and 157 $\rm \mu m$ are shown in Figs. \ref{LineSpec63} and \ref{LineSpec157} respectively.

\begin{table}[!hd]
\centering
\caption{\textit{Herschel}/PACS spectroscopy}             
\label{HSOspec}              
\begin{tabular}{lllllll}     % 6 columns 
\hline\hline       
Name & [OI] flux & S/N & [CII] flux & S/N\\ 
 &  & $\rm 63 \mu m$ & & $\rm 157 \mu m$ \\
--	& $\rm (10^{-18} W/m^{2})$ & --  & $\rm (10^{-18} W/m^{2})$ & --\\ 
\hline     
HD 164249 & $\rm < 8.2 $ & 2.6 & $\rm < 3.6$ & 0.3 \\
HD 172555$\rm ^{1}$ & $\rm 9.7 \pm 2.0$ & 3.0 & $\rm < 2.5$ & 1.5 \\
HD 181296 & $\rm < 6.2 $ & 3.6 & $\rm 2.3 \pm 0.6$ & 3.2\\
HD 181327 & $\rm < 8.2 $ & 10.4 & $\rm < 7.6$ & 5.8 \\
\hline                  
\end{tabular}
\tablefoot{Columns are target name, [OI] line flux at 63.18 $\rm \mu m$, S/N of the continuum at 63 $\rm \mu m$, [CII] line flux at 157.74 $\rm \mu m$, and S/N of the continuum at 157 $\rm \mu m$. (1): Recentred and averaged spectrum.}
\end{table}

The presence of atomic oxygen in \object{HD~172555} was discussed in \cite{Riviere2012}, and we review the main results in Section \ref{GasBPMG}. HD 181296 shows [CII] emission at 157.74 $\rm \mu m$ with a S/N of 3.8. This is the first detection of gas emission toward HD 181296, adding HD 181296 to the short list of debris disc with a gas detection. We discuss the detection in detail in Section \ref{GasBPMG}. None of the systems show water emission at 63.32 $\mu \rm{m}$.

\section{Blackbody models}\label{dustModels_Sec}

\begin{figure*}[!t]
\begin{center}
   \centering
     \includegraphics[scale=0.35, trim = 0mm 0mm 0mm 0mm,clip]{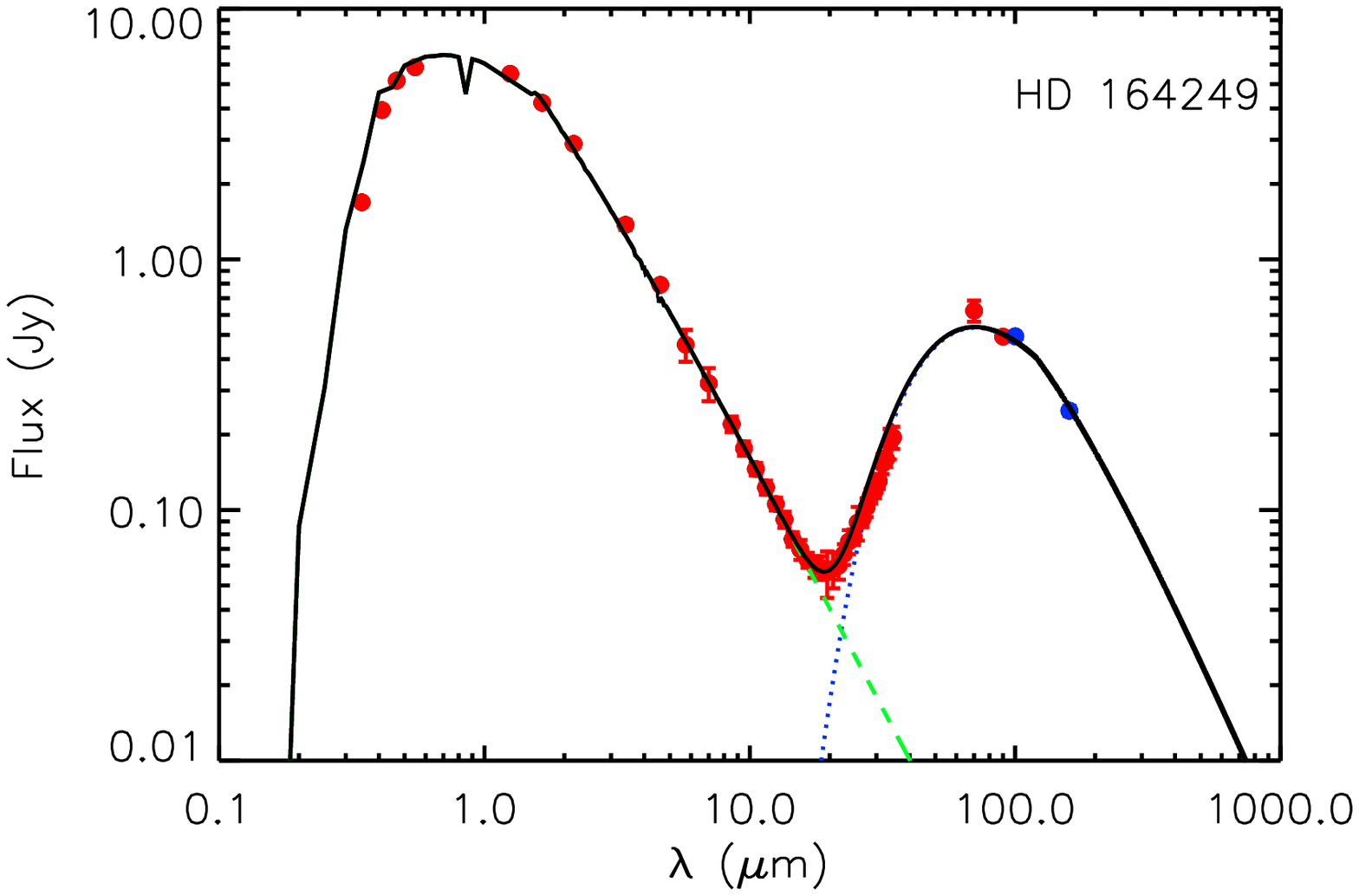}       
     \includegraphics[scale=0.35, trim = 13mm 0mm 0mm 0mm,clip]{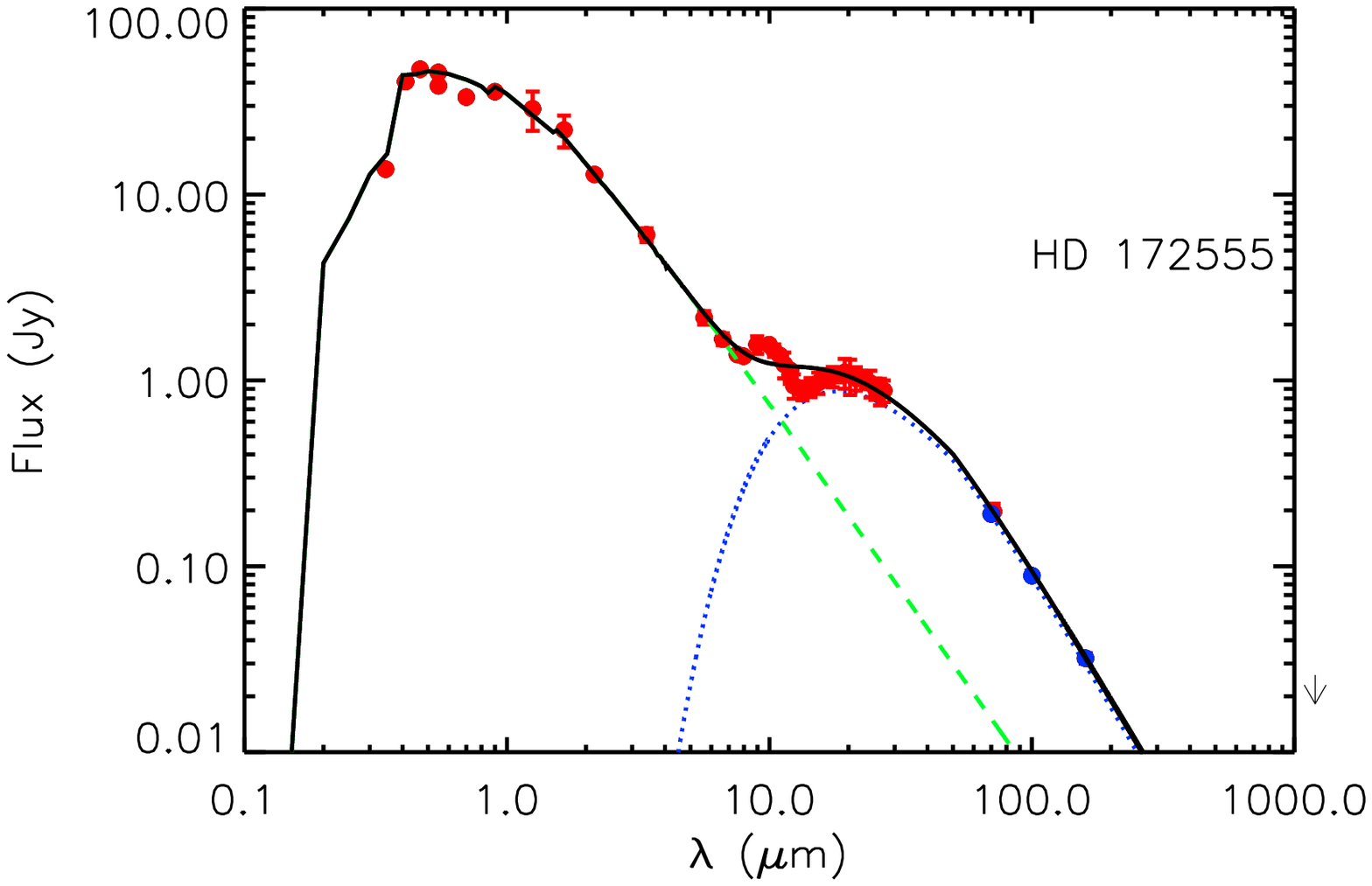}  
     \includegraphics[scale=0.355, trim = 13mm 0mm 0mm 0mm,clip]{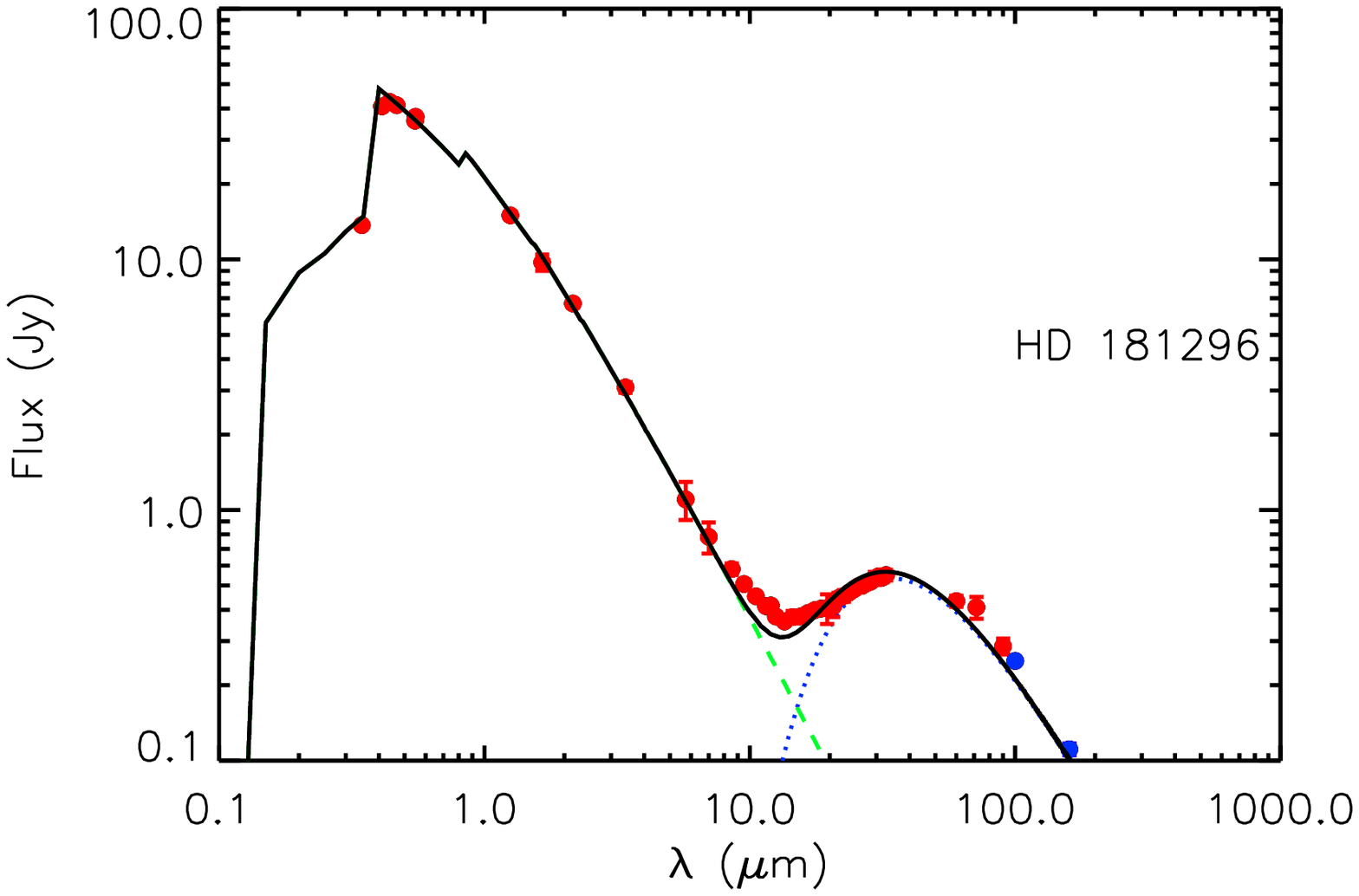}\\
     \includegraphics[scale=0.35, trim = 0mm 0mm 0mm 0mm,clip]{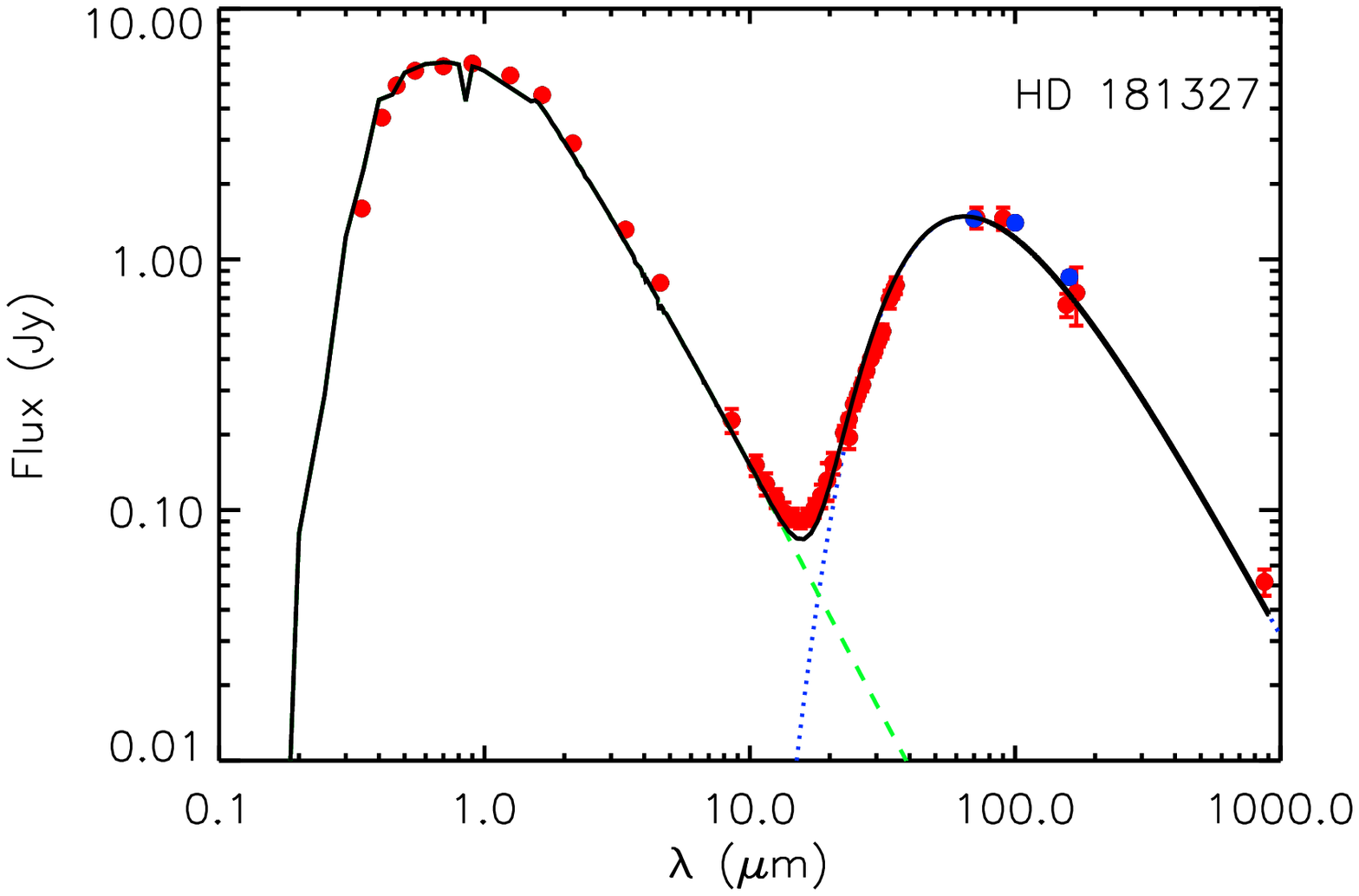}     
     \includegraphics[scale=0.35, trim = 13mm 0mm 0mm 0mm,clip]{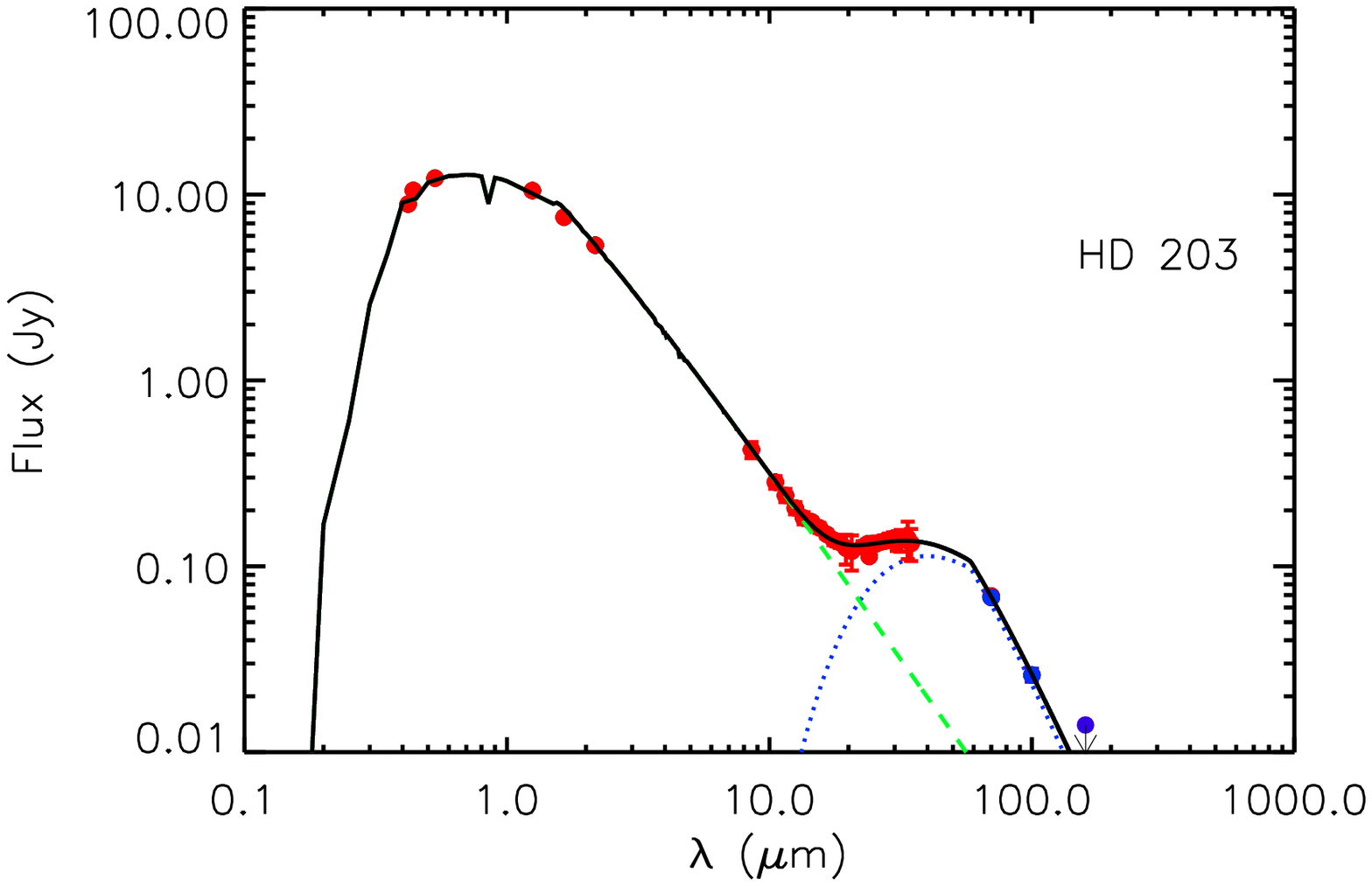}     
     \includegraphics[scale=0.35, trim = 13mm 0mm 0mm 0mm,clip]{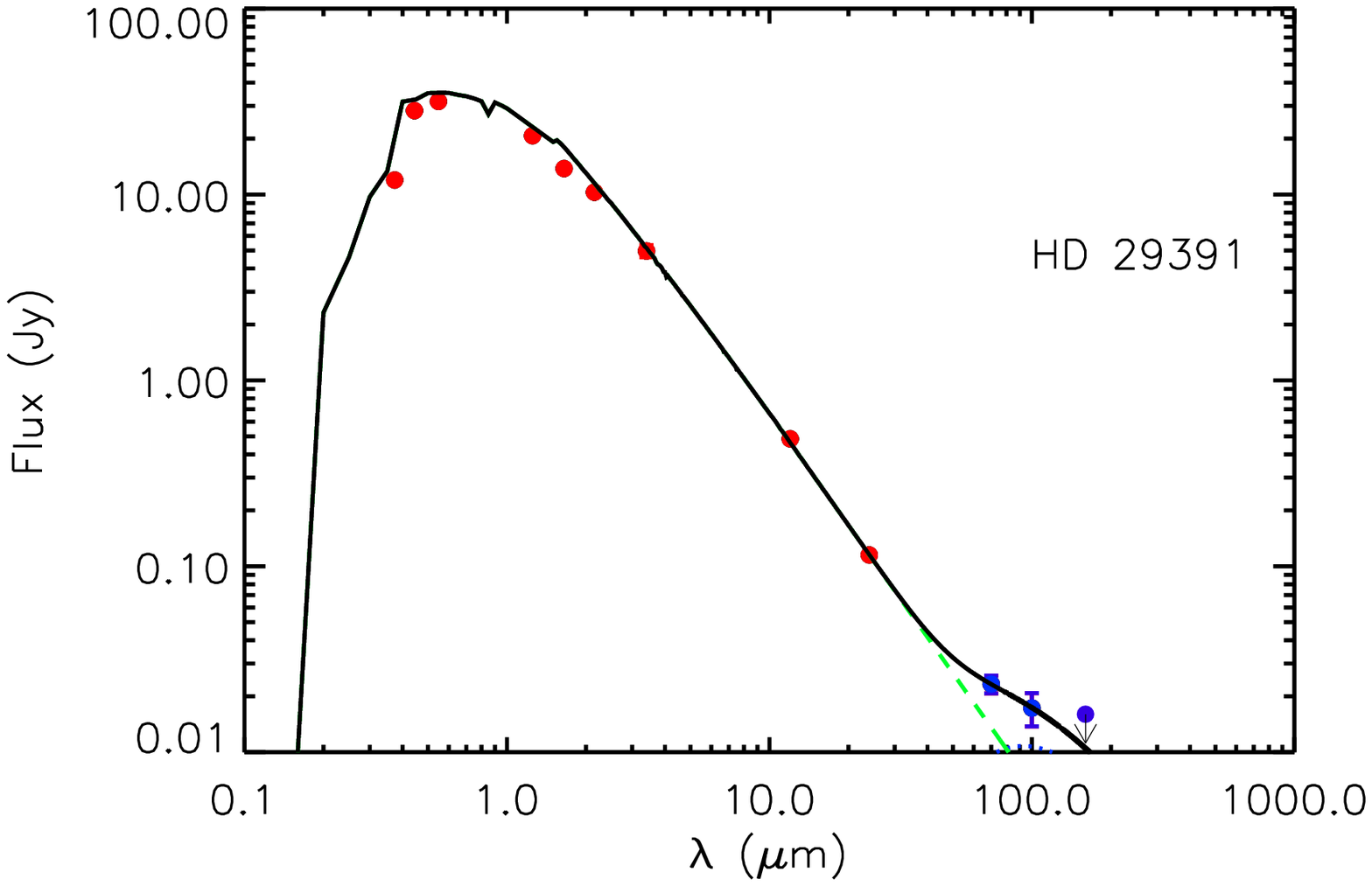} \\
     \includegraphics[scale=0.35, trim = 0mm 0mm 0mm 0mm,clip]{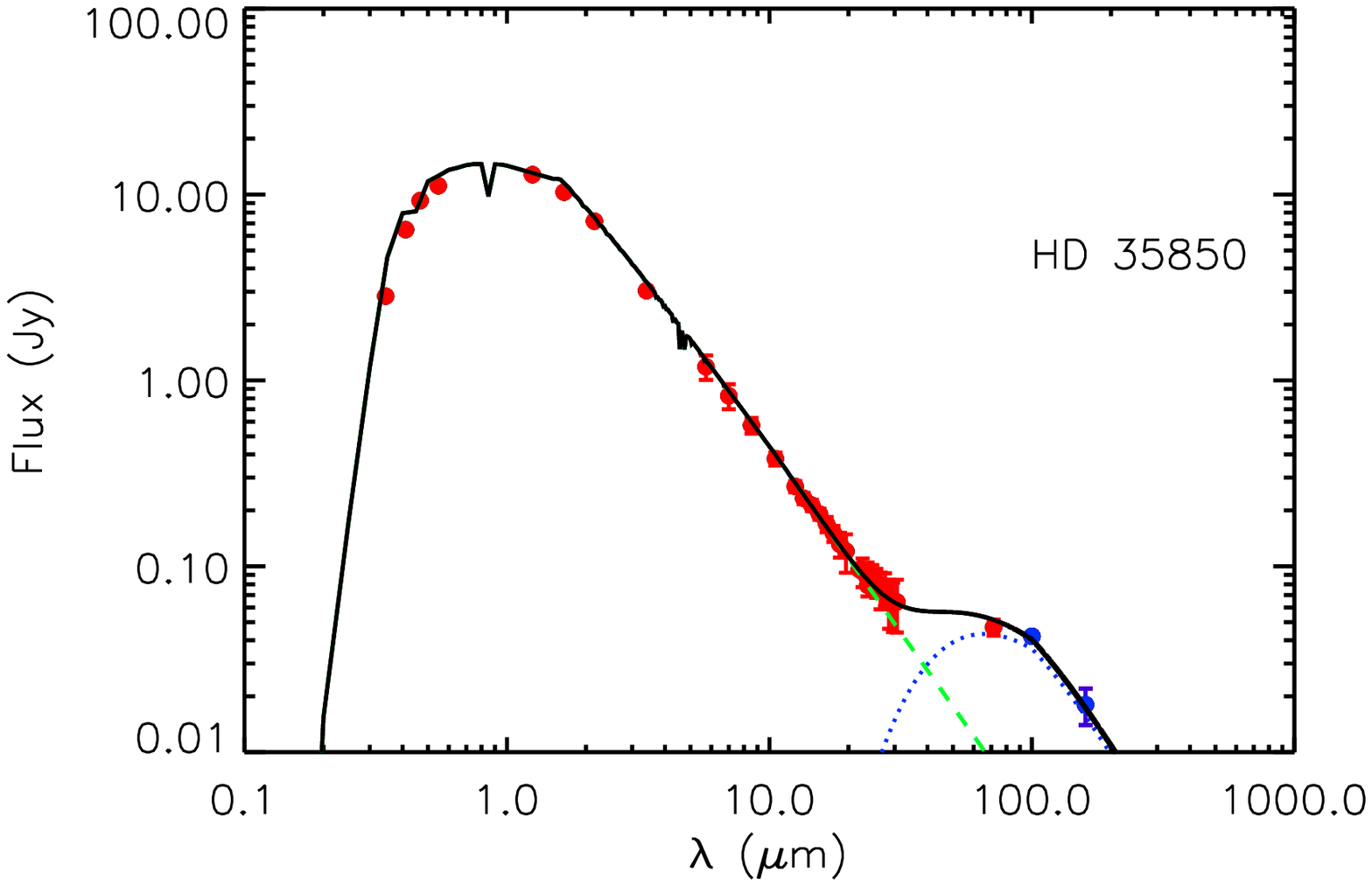}     
     \includegraphics[scale=0.35, trim = 13mm 0mm 0mm 0mm,clip]{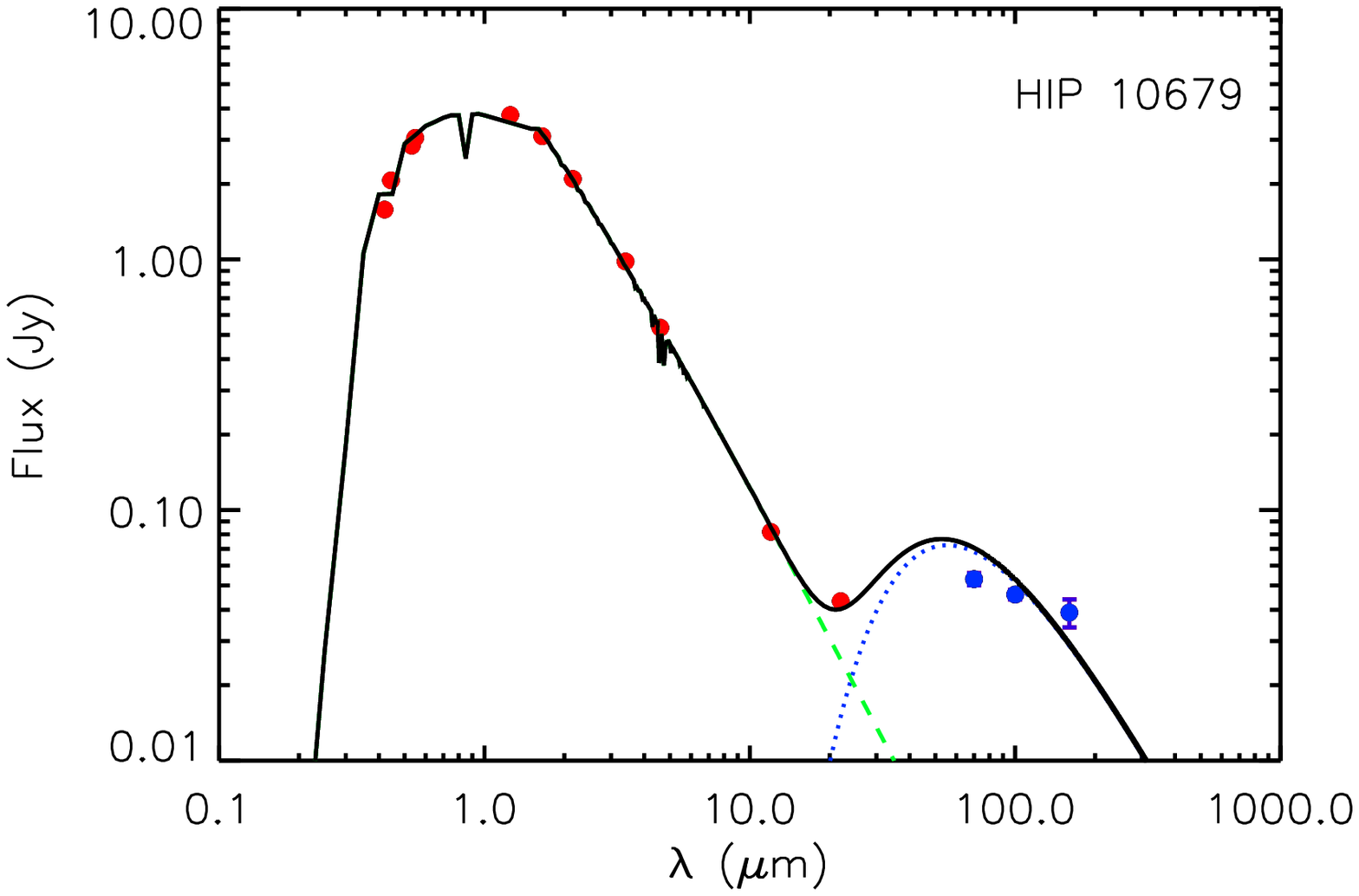} 
     \includegraphics[scale=0.35, trim = 13mm 0mm 0mm 0mm,clip]{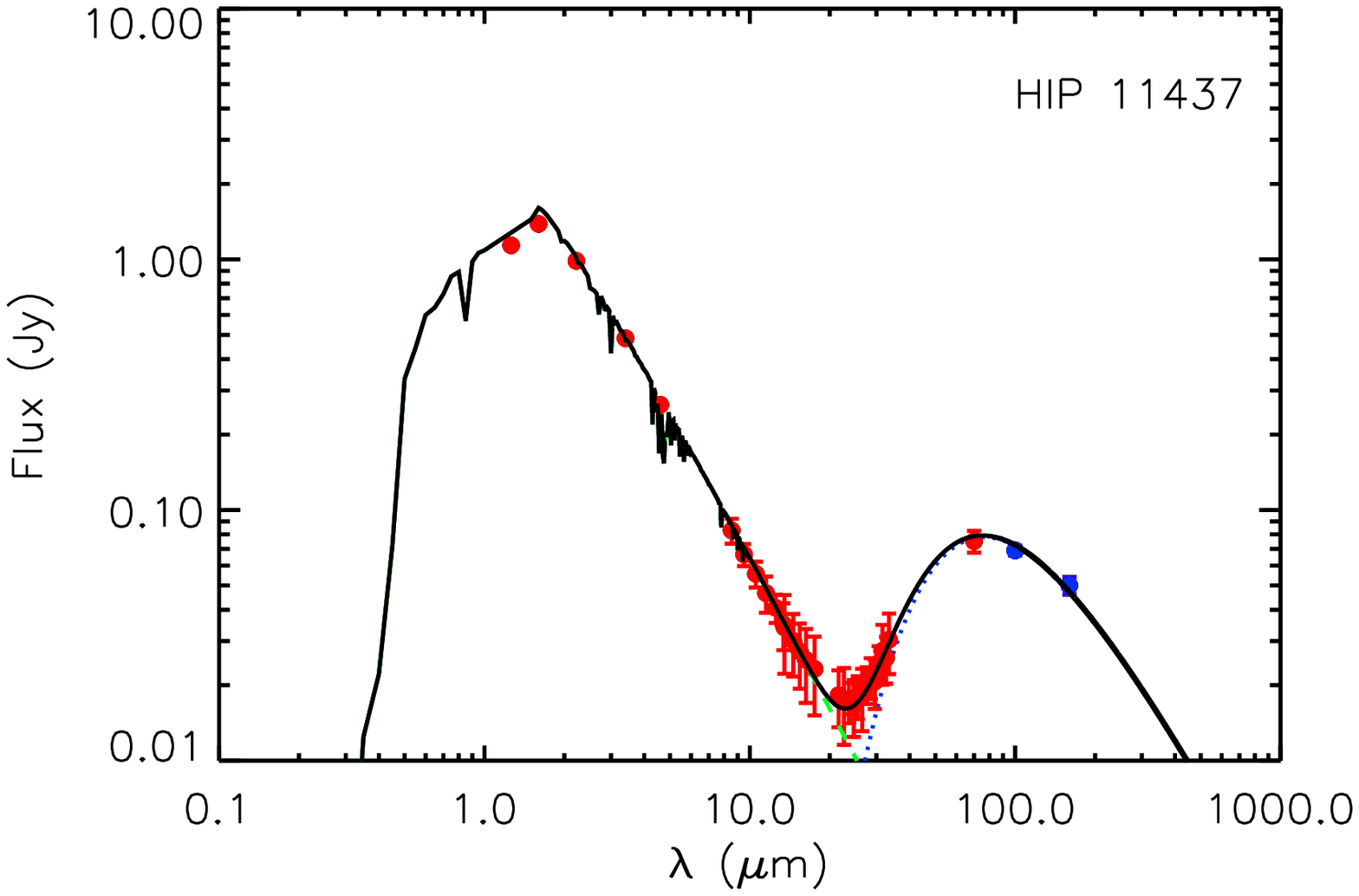}
     \caption{Blackbody models for BPMG members detected with PACS. Red dots are photometric points from the catalogues and literature (see Sec. \ref{BPMG:sample}). Blue dots are PACS observations from this work. The black solid line represents the photosphere plus blackbody model. The doted blue line represents the blackbody that better fits the dust. The green dashed line represents the photospheric contribution.}
   \label{BBmods}
\end{center}
\end{figure*}

For each BPMG member in the sample, we have built the SED and included the new PACS photometric data, aiming to compare the observed photometry with synthetic models to study the BPMG disc properties. The simplest and most  robust way of studying debris disc properties in BPMG members is to fit modified blackbody models to the dust emission. The photospheric contribution is estimated using photospheric models described by the stellar parameters listed in Table~\ref{tableStar}, with solar metallicity and $\rm{log}g=4.5$. The modified blackbody is described by	
\begin{equation}
{\rm F \propto F_{\nu}(T_{dust}) \times (\lambda_{0}/\lambda)^{\beta}}
\end{equation}
where $\rm F_{\nu}$ is the flux emitted by a blackbody at a temperature $\rm T_{dust}$, $\rm \beta =0$ if $\rm \lambda < \lambda _{0}$. The parameters $\rm \beta$, $\rm \lambda_0$ and $\rm T_{dust}$ are free to vary. Here, $\rm \lambda_{0}$ can be interpreted as a characteristic grain size, via $\rm \lambda_{0}=2\pi a$ where a is the average grain size.

Such simple models provide estimates of the dust temperature and IR excess. The best fit model was determined through a reduced $\rm \chi^{2}$ minimisation, with
\begin{equation}\label{EqChi}
 \chi^{2}_{\nu}={1 \over \nu} \sum_{i=1}^{N} {(F_{obs,i}-F_{mod,i})^{2} \over \sigma_{i}^{2} }
\end{equation}
where N is the total number of photometric points used in the fitting, $\rm \nu = N-n$ the degrees of freedom, n the number of free parameters, $F_{obs,i}$ the observed flux of the \textit{i}-th photometric point, $F_{mod,i}$ the  \textit{i}-th model flux, and $\rm \sigma$ is the error of the \textit{i}-th photometric point. Because the stellar properties are fixed, we only included excess points in the analysis. For the sources with IRS data available, we used the IRS spectra turning point as the starting point for the excess. When no IRS spectra were available, we just used the first photometric point showing excess. 

We used a genetic algorithm (\textit{GAbox}\footnote{This algorithm has already been used in \cite{Riviere2013} with similar purposes and is explained in Appendix A of \cite{lillo-box14}}) to explore the parameter space and find the set of free parameters that minimises Eq.~\ref{EqChi}. \textit{GAbox} allowed us to rapidly identify the set of parameters that best fits our observations. Then, it conducts the estimation of the uncertainties by creating a small grid of parameters around the best solution and computes the $3\sigma$ confidence levels for each parameter pair. The largest uncertainty among all pairs involving a particular parameter is then selected as the final uncertainty.

From the dust temperature of the blackbody models, the lower limit on the disc radius can be estimated by using  
\begin{equation}\label{eq:Radius}
R_{\rm in} > \frac{1}{2} R_{*}\left(T_{*} \over T_{\rm
  dust}\right)^{2}
\end{equation}
where $\rm T_{*}$ and $\rm R_{*}$ are the star effective temperature and radius. We note here that inner radii derived from blackbody dust temperatures are lower limits to the actual disc radii \citep[see e. g.][]{Rodriguez2012}. The dust mass can be estimated using
\begin{equation}\label{eq:dustMass}
M_{dust}={F_{\nu}(\lambda_{Obs})D^{2} \over \kappa_{\nu} B_{\nu}(T_{dust})}
\end{equation}
where $D$ is the
distance to the star, $\rm{B}_{\nu}(\rm{T_{dust}})$ can be approximated by
the Rayleigh-Jeans regime, $\kappa_{\nu}\rm{=2\times (1.3 mm /}
\lambda)^{\beta}~\rm{cm^{2}g^{-1}}$ \citep{Beckwith1990}, and $ \rm{F}_{\nu}(\lambda_{Obs})$ is the observed integrated flux density at a given wavelength emitting in the Rayleigh-Jeans regime. We use the flux  at 160 $\rm \mu m$, since dust emission is always in the Rayleigh-Jeans regime for BPMG members at this wavelength \citep[see][]{Harvey2012,Riviere2013}.
\begin{table*}[!t]
\caption{Black body models}
\label{BBmodels}
\centering
\begin{tabular}{lllllll}   
\hline
\hline       
Name & T & $\beta$ & $\rm \lambda_0$&$\rm L_{IR}/L_{*}$ & $\rm R_{dust}$ & $\rm M_{dust}$\\ 
 -- & (K) & -- & ($\rm \mu m$) & -- & (AU) & ($\rm M_{\oplus}$) \\ 
\hline 
\object{HD~203} & $\rm 128^{+10}_{-11} $ & $\rm 1.6^{+0.56}_{-0.50}$ & $\rm 58.4^{+8.0}_{-8.3} $ &  $\rm 1.6 \times 10^{-4}$ & $\rm 9.1 \pm 3.0$ & $\rm < 6.6 \times 10^{-5}$ \\
\object{HD~29391} &  $\rm 55^{+21}_{-21}$ & $\rm 0.0^{+2.0}$ &  $\rm 9.5^{+111} $ &$\rm 2.3 \times 10^{-6}$ & $\rm 82^{+677}_{-75}$ & $\rm < 1.6 \times 10^{-3}$\\ 
\object{HD~35850} &  $\rm 77.6^{+8.5}_{-15}$& $\rm 0.7^{+1.3}_{-0.7}$ & $\rm 100^{+60}_{-98}$  & $\rm 3.7 \times 10^{-5}$ & $\rm 16.7^{+10.3}_{-7.5}$ & $\rm (3.4 \pm 1.4) \times 10^{-4}$ \\
\object{HD~164249} &  $\rm 71.15^{+0.99}_{-0.54}$  & $\rm 1.16^{+0.84}_{-0.90}$ & $\rm 119 \pm 6 $ & $\rm 8.4 \times 10^{-4}$ & $\rm 25.2 \pm 4.6$ & $\rm (6.8^{+15.7}_{-5.3}) \times 10^{-3}$\\
\object{HD~172555} & $\rm 264.2 ^{+8.5}_{-8.4}$ & $\rm 0.45^{+0.19}_{-0.11} $& $\rm 29.9^{+11}_{-8.9}$ & $\rm 7.5 \times 10^{-4}$ & $\rm 2.76 \pm 0.77$ & $\rm (2.7 \pm 1.2) \times 10^{-4}$\\
\object{HD~181296} & $\rm 161.7^{+2.9}_{-3.0}$& $\rm 0.0^{+2.0}$ & ($\rm 160_{-47}$) & $\rm 2.4 \times 10^{-4}$ & $\rm 17.0 \pm 1.8$ & $\rm (1.3 \pm 0.08) \times 10^{-2}$ \\ 
\object{HD~181327} & $\rm 79.04^{+0.07}_{-0.33}$ & $\rm 0.0^{+0.20}$ & ($\rm 311^{+353}_{-290}$) &$\rm 2.9 \times 10^{-3}$ & $\rm 22.6 \pm 2.9$ &$\rm  (0.20 \pm 0.06) $\\
\object{HIP~10679} &$\rm 96.2^{+2.9}_{-3.1}$ & $\rm 0.0^{+2.0}$ & ($\rm 160_{-147}$) & $\rm 3.0 \times 10^{-4} $ & $\rm 8.7 \pm 2.5$ & $\rm (3.7 \pm 0.6) \times 10^{-3}$\\ 
\object{HIP~11437} & $\rm 65.5^{+22}_{-2.9}$ & $\rm 0.1^{+1.9}_{-0.1}$ & ($\rm 18^{+179}_{-15}$) & $\rm 1.0 \times 10^{-3}$ & $\rm 8.3 \pm 1.7$ & $\rm (9.8 \pm 0.9) \times 10^{-3}$ \\
\hline
\object{AT~Mic} & 20 & 0.0 & -- & $\rm < 2.5 \times 10^{-5}$ & -- & -- \\ 
\object{CD~64-1208} & 70 & 0.0 & -- &  $\rm < 4.2 \times 10^{-5}$ & -- & -- \\ 
\object{GJ~3305} & 70 & 0.0 & -- &  $\rm < 4.5 \times 10^{-5}$ & -- & -- \\ 
\object{HD~45081} & 70 & 0.0 &-- &   $\rm < 9.4 \times 10^{-6}$ & -- & -- \\ 
\object{HD~139084} & 70 & 0.0 & -- &  $\rm < 8.4 \times 10^{-6}$ & -- & -- \\ 
\object{HD~146624} & 70 & 0.0 & -- &  $\rm < 5.0 \times 10^{-7}$ & -- & -- \\ 
\object{HD~174429} & 70 & 0.0 & -- &  $\rm < 2.3 \times 10^{-5}$ & -- & -- \\ 
\object{HD~199143} & 70 & 0.0 & -- &  $\rm < 2.6 \times 10^{-6}$ & -- & -- \\ 
\object{HIP~10680} & 70 & 0.0 & -- &  $\rm < 8.1 \times 10^{-6}$ & -- & -- \\ 
\object{HIP~12545} & 70  & 0.0 & -- &  $\rm < 4.2 \times 10^{-5}$ & --& -- \\ 
\hline
\end{tabular}
\tablefoot {For objects without a detected excess, we fixed $\rm \beta = 0.0$.}
\end{table*}

For the sources without detected excess, we computed upper limits on the infrared excess luminosity by using the upper limit with the smaller excess over the photosphere to scale the flux of a 70 K pure blackbody ($\rm \beta =0$). The temperature was chosen to be similar to the median temperature of the objects with excess. Temperatures higher than 70 K can result in an excess over the photosphere at $\rm \lambda \sim 30~\mu m$ where the IRS spectra do not show any excess. For AT Mic, the detection of the photosphere at 70 $\rm \mu m$ makes the blackbody model fitting more restrictive: models with $\rm T_{dust} \ge 25 K$ result in an excess flux at 70 $\rm \mu m$ that is not observed, therefore we fixed the dust temperature to be 20 K. Results of the fitting are listed in Table~\ref{BBmodels}, and model SEDs can be found in Fig. \ref{BBmods}. 

\section{Discussion} 

\begin{figure}[!b]
\begin{center}
   \centering
     \includegraphics[scale=0.55,trim= 12mm -3mm 0mm 0mm,clip]{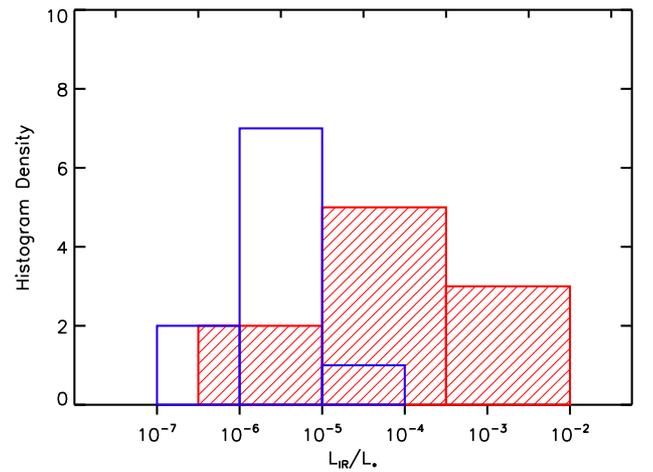} 
     \caption{Histograms with the distribution of the infrared fractional excess for objects detected (red histogram filled with lines) and non-detected (blue histogram) with PACS.}
   \label{LIRhist}
\end{center}
\end{figure}

In Fig. \ref{LIRhist} we show an histogram of the infrared fractional excess ($\rm L_{IR}/L_{*}$) distribution of BPMG members observed with PACS, including $\beta $ Pic data from \cite{Rebull2008}. The infrared fractional excess is computed from blackbody models as the fraction of dust luminosity over the stellar luminosity. The distribution of detected sources peaks at $\rm \sim10^{-4}$, while the distribution of non-detected sources upper limits peaks at $\rm \sim10^{-6}$. However, the faint tail of the distribution of detected objects coincides with the peak of the distribution for non-detected objects. The IR excess in BPMG members is due to the presence of a debris disc, with dust temperatures in the range 55 to 264 K, and a median temperature of 79 K. In the BPMG sample of objects observed with PACS, six out of nine objects have temperatures in the range 55 to 96 K and are therefore Kuiper belt-type discs. However, three out of nine, namely HD 203, HD 172555, and HD 181296, show temperatures that are much higher and may consist of a warm component or both a warm and a cold component, such as the one found in $\rm \beta$ Pic itself. Previous work suggests that a significant fraction of debris discs might have two-component architecture \citep[see e. g.][]{Morales2009,Su2013}. Derived dust masses range from $\rm < 6.1 \times 10^{-5}~M_{\oplus}$ to  $\rm 0.20~M_{\oplus}$ and have an inner radii from 2.0 to 60 AU.

The detection ratio of objects with an IR-excess is $\rm \sim 50Ê\%$, much higher than the standard value of 15\% to 20 \% for older systems \citep{Beichman2006,Bryden2006}. A trend towards a higher detection rate of debris discs among young stars is noted by \cite{Rieke2005}. However, our ratio is computed using a small sample that is not robust against selection effects.

\subsection{Notes on the individual SEDs}
\object{HD~203} was not detected at 160 $\mu \rm{m}$. The best fit temperature was $\rm T_{dust}=128~K$, in good agreement with \cite{Rebull2008}. The way far-IR emission plummets around 160 $\mu \rm{m}$ suggests an under-abundance of large grains. An unusual, steep decrease in the SED in the PACS wavelength range of three debris discs in the DUNES survey is reported by \cite{Ertel2012}, who define
\begin{equation}
\Delta_{\nu_{1},\nu_{2}} ={log~F_{\nu_{2}}-log~F_{\nu_{1}} \over log~\nu_{2} - log~\nu_{1}}
\end{equation}
and find $\Delta_{70,100}$ values in the range 1.94 to 2.66, while we get $\Delta_{70,100} = 2.70$ for HD 203. The authors consider that the steep decay of the SED is the result of a deviation from the standard equilibrium in a collisional cascade, with a significative under-abundance of large grains. A similar SED decay in the PACS wavelength range is also reported by \cite{Donaldson2012} for \object{HD~3003}, where a value of -4.4 was used for the exponent of the grain size distribution. Overall, it seems that we need a grain size distribution that is different from the expected in a collisional cascade in equilibrium to explain steep SEDs.

Our observations of HD 29391 show for the first time that the star is surrounded by a debris disc that produces a very tiny excess ($\rm L_{IR}/L_{*}=2.0\times 10^{-6}$, the faintest in the sample), at a temperature of only 55 K, the coldest disc detected in BPMG. The lower limit on the inner radius is 82 AU, the largest among the sample, but it is poorly constrained due to large errors in $\rm T_{dust}$. We computed an upper limit in dust mass of $\rm M_{dust}<1.6 \times 10^{-3}$.

\object{HD~172555} is an A7 star that harbours a warm debris disc with solid silicate feature emission in its IRS spectrum \citep{Chen2006} and with an infrared fractional luminosity that is $\sim 86 $ times higher than the maximum value expected from steady-state collisional evolution \citep{Wyatt2007}. \cite{Lisse2009} analysed its IRS spectrum and propose that both an SiO gas tentative detection and a silicate feature around 10 $\mu \rm{m}$ could be explained as the outcome of a hypervelocity collision ($\rm > 10~ km~s^{-1}$) between two planetary mass objects. The best modified blackbody model for \object{HD~172555} has $\rm T_{dust}=264~K$ ($\rm \beta = 0.5$), the warmest in the sample. Although it is a poor fit for the IRS data, it demonstrates there is warm dust in the disc.After re-analyzing TReCs observations \citep{Moerchen2010}, \cite{Smith2012} show that the disc emission is resolved in the Q band ($\rm 18.3 ~Ê\mu m$), arising from $\rm r > 8 ~AU$,  but not in the N band ($\rm 11.66 ~\mu m$), with the emission coming from 1.0 to 7.9 AU from the central star. These results suggest that there may be two dust populations in the system.

The SED of HD 181327, including the PACS data from GASPS, has recently been modelled by \cite{Lebreton2012}, taking the imaging constraints by \cite{Schneider2006} into account, with a dust mass of $\rm 0.05~M_{\oplus}$ in grains from 1 to $\rm 1000~\mu m$ composed of a mixture of silicates, carbonaceous material and amorphous ice with significant porosity. The best fit temperature for HD 181327 is 79 K, which results in a lower limit for the disc radius of $\rm \sim 23~AU$, with a dust mass of $\rm 0.20~M_{\oplus}$. We note here that the minimum radius of the disc is well below the value of 86 AU by \cite{Schneider2006}, obtained by means of NICMOS coronographic observations of scattered light. That blackbody radii are a lower limit to actual debris disc radii is a well known fact \citep[see e. g. ][]{Booth2013}.

The IRS spectrum for HIP 10679 was too noisy to be included in the analysis of the SED. In Fig. \ref{BBmods} we can see that with a single blackbody fit we can not reproduce its IR emission. The best fit model, with $\rm T = (92 \pm 4) ~K$ and $\rm \beta=0.0$, produces a poor fit that clearly underestimates the flux at 160 $\mu \rm{m}$, while overestimating the flux at 70 $\mu \rm{m}$, arguing for a flatter slope. Our best fit temperature results in an inner radius of  $\rm (8.7 \pm 2.5)~AU$, much smaller than the value of 35 AU proposed by \cite{Rebull2008}. 

In Fig. \ref{BBmods} we can see that a single blackbody  also fails to reproduce the IR emission towards HD 181296. The best fit model, with T = $\rm 161 ~K$ and $\rm \beta=0.0$ underestimates the flux at far-IR wavelengths. The poor fit quality might be explained by the need for a second, warm blackbody. The system was imaged by \cite{Smith2009}, who derived $\rm R_{out}=45~AU$ and propose that two different grain populations should be present.

\subsection{Gas in BPMG debris discs}\label{GasBPMG}

There are three circumstellar systems in the BPMG with detected gas emission: \object{$\rm \beta$ Pic}, \object{HD~172555}, and \object{HD~181296}. Given the age of the BPMG, a primordial origin for the gas is unlikely. The disc around $\rm \beta$ Pic has long been known to possess large amounts of gas in its disc \citep{Hobbs1985,Lagrange1986,Roberge2000,Olofsson2001,Thi2001}. \cite{Lagrange1987} showed that UV absorption lines profiles in $\rm \beta$ Pic show variations on time scales of months, signature of infalling gas \citep{Lagrange1989}. \cite{Olofsson2001} show Na I gas in Keplerian rotation. A braking mechanism is needed to keep Na stable against radiation pressure; \cite{Fernandez2006} propos that an overabundance of C by a factor of 10 with respect to solar abundance is enough to brake the gas. Observations with the Far-UV Space Explorer (FUSE) have shown that C is indeed overabundant by a factor of 20 with respect to other species \citep{Roberge2006}.

The detection of atomic oxygen emission at 63.18 $\mu \rm{m}$ towards HD 172555 has previously been reported in \cite{Riviere2012b}, and therefore we refer the reader to this paper, where a possible origin for the emission in a hypervelocity collision is discussed \citep[see also][]{Lisse2009, Johnson2012}. \cite{Lisse2009} propos a tentative detection of SiO gas, later on confirmed by \cite{Johnson2012}, who concluded that the amount of oxygen gas proposed by \cite{Riviere2012b} is enough to keep SiO vapour from being destroyed by photo-dissociating photons. More recently, \cite{Kiefer2014} reported the detection of variable absorption signatures in the Ca II H and K lines, a signature of falling evaporating bodies (FEB), highlighting the similarities between $\beta$ Pic and HD 172555.

\begin{figure}[!ht]
\begin{center}
   \centering
     \includegraphics[scale=0.5,trim= 0mm 0mm 0mm 0mm,clip]{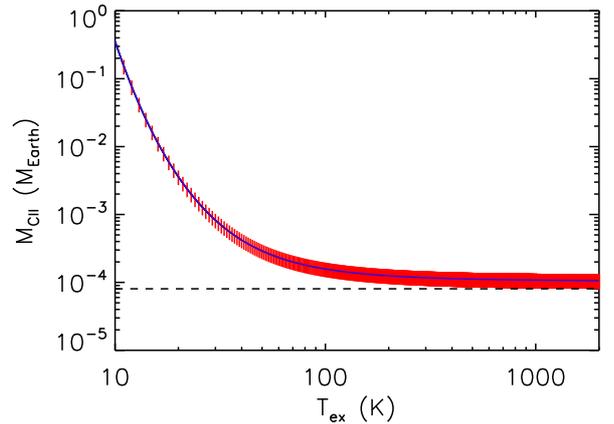} 
     \caption{Gas mass versus excitation temperature for [CII] in HD 181296. The red, vertical bars show the error in the [CII] gas mass determination. The doted, horizontal bar depicts the position of the lower limit to the [CII] gas mass.}
   \label{HD181296_CII_gas}
\end{center}
\end{figure}

We report the first detection of [CII] emission at 157 $\rm \mu m$ towards HD 181296, with a flux of $\rm (1.7 \pm 0.4) 10^{-18} W/m^{2}$. This is the third case of a debris disc where [CII] emission is present but no [OI] emission is observed, the other two systems being \object{HD~32297} \citep{Donaldson2013} and \object{49~Cet} \citep{Roberge2013}. Following \cite{Roberge2013}, and assuming the [CII] emission is optically thin, we can use
\begin{equation}
M_{[CII]}={4 \pi \lambda_{[CII]} \over hc}{F_{int} m d^{2} \over A_{10} \chi_{u}}
\end{equation}
to estimate a lower limit for the [CII] gas mass, where $\lambda_{[CII]}$ is the rest frame wavelength of the transition, $F_{int}$ the integrated line flux, $m$ the mass of an atom, $d$ the distance to the source, $A_{10}=2.4 \times 10^{-6} ~s^{-1}$ the spontaneous transition probability, and $\chi_{u}$ the fraction of atoms in the upper level. Assuming local thermal equilibrium (LTE), we can compute $\chi_{u}$ using
\begin{equation}
\chi_{u}={(2J_{u}+1) \over Q(T_{ex})} e^{-E_{ul} / KT_{ex}} 
\end{equation}
where $J_{u}$ is the angular momentum quantum number of the upper level,  $g_{u}$ the statistical weight of the upper level, $E_{ul}$ the energy difference between the upper and the lower levelS ($E_{1}/k=91.21~K$), $T_{ex}$ the excitation temperature and $Q(T_{ex})$ the partition function, which depends on the excitation temperature. Assuming a two-level atom, $Q(T_{ex})$ is
\begin{equation}
Q(T_{ex})=g_{l} + g_{u}e^{-E_{ul} / kT_{ex}}
\end{equation}
where $g_{l}$ is the statistical weight of the lower level. In Fig. \ref{HD181296_CII_gas} we show the distribution of $\rm M_{[CII]}$ as a function of $\rm T_{ex}$. Different excitation temperatures have led to very different gas masses, but we computed a lower limit of $\rm 8.1 \times 10^{-5}~M_{\oplus}$ valid in the range 1--2000 K. To translate this [CII] mass lower limit into a total C mass lower limit, we need knowledge about the ionization fraction. Following \cite{Roberge2013}, we assume that HD 181296 has the same ionization fraction as \object{$\beta$~Pic} ($\rm 50\%$), given that they have a similar spectral type, both have optically thin debris discs, and assume that the ionization fraction is similar for both infrared fractional luminosities. A lower limit to the total carbon mass is then $1.6 \times 10^{-4}~M_{\oplus}$. If we assume that the carbon has a solar abundance (solar carbon mass fraction is 0.288\%), we get $\rm M_{gas} \ge 0.056 ~M_{\oplus}$.

The origin of the gas in HD 181296 is difficult to assess without more species detected. The most probable mechanisms generating secondary gas in debris discs are grain-grain collisions, sublimation of dust grains, evaporation of comets, and photo-desorption from dust grain surfaces. Photo-desorption of CO from grain surfaces, followed by photo-dissociation, should leave a similar amount of C and O atomic gas in the system, but we do not detect any [OI] emission. Therefore, it is unlikely that photo-desorption of CO is the main production mechanism. However, \cite{Roberge2000,Roberge2006} show that in the case of $\beta$ Pictoris only 2\% of carbon gas could come from dissociation of CO, implying that many other materials could be acting as [CII] suppliers, such as amorphous carbon.

\begin{table}[!hd]
\centering
\caption{Overview of gas detections in debris discs. }             
\label{DDgas}              
\begin{tabular}{lcccc}     % 6 columns 
\hline\hline       
Name & Sp. type  & Age & $\rm T_{dust}$& Reference \\ 
-- & -- & (Myr) & (K)  & --\\ 
\hline     
\object{$\rm~\beta$} Pic & A6$\rm ^{1}$ & $\rm 20 \pm 10$ & 130$\rm ^{2}$ & 3 \\
\object{$\sigma~Her$} & B9$\rm ^{4}$ & 200$\rm ^{5}$ & 200$\rm ^{6}$ & 4 \\
\object{HD~21997} & A3$\rm ^{7}$ & $\rm 20 \pm 10$$\rm ^{8}$ & 52$\rm ^{9}$ & 10 \\
\object{HD~32297} & A0$\rm ^{11}$ & 30$\rm ^{12}$ & 240$\rm ^{13}$ & 13 \\
\object{HD~172555} & A6$\rm ^{11}$ &  $\rm 20 \pm 10$ & 264 & 14 \\
\object{HD~181296} & A0$\rm ^{11}$ & $\rm 20 \pm 10$ & 162 & 15 \\
\object{49~Cet} & A1$\rm ^{16}$ & $\rm 15-180$$\rm ^{17}$ & 175$\rm ^{16}$ & 16 \\
\object{51~Oph} & B9.5$\rm ^{18}$ & 0.2-1.1$\rm ^{17,*}$ & 400--1000$\rm ^{19}$  & 20\\
\hline                  
\end{tabular}
\tablefoot{$\rm T_{dust}$ refers to the temperature of the warm dust when two components are present.\\
(1): \cite{Gray2006}. (2): \cite{Rebull2008}. (3): \cite{Vidal-Madjar1998}. (4): \cite{Chen2003}. (5): \cite{Grosbol1978}. (6): \cite{Fajardo1998}. (7): \cite{Houk1988}. (8): \cite{Moor2006}. (9): \cite{Nilsson2010}. (10): \cite{Moor2011}. (11): \cite{Torres2006}. (12): \cite{Kalas2005}. (13): \cite{Donaldson2013}. (14): \cite{Riviere2012}. (15): present work. (16): \cite{Roberge2013}. (17): \cite{Montesinos2009}. (18): \cite{Thi2013}. (19): \cite{Fajardo1993}. (20): \cite{Lecavelier1997}.\\
(*): \cite{Thi2013} consider it is a debris disc despite its young age.}
\end{table}

It is interesting to note that the presence of gas might be connected with the presence of a warm disc: three out of three BPMG members with a gaseous debris discs also have a hot excess component. In Table~\ref{DDgas} we list debris discs with confirmed gas detections. We do not include the four shell stars proposed by \cite{Roberge2008}  in the list nor \object{TWA~04B} \citep{Riviere2013} or \object{HD~141569A} \citep{Thi2013}, because their evolutionary stages need to be clarified. Seven out of eight debris discs with a solid gas detection have a warm disc component, the only exception being HD 21997, with a dust temperature of 52 K \citep{Nilsson2010}. However, \cite{Kospal2013} argue that the gas surrounding HD 21997 could be a remnant of primordial gas. While the connection between the presence of hot dust and that of gas requires further research, the significance of this trend is tantalizing. Considering this, HD 203 is a good candidate for future gas observations.

Another interesting fact is that the three stars share similar spectral spectral type (A0, for HD 181296, to A6 for HD 172555 and $\beta$ Pic), which could imply that photo-desorption of gas from grain surfaces plays a major role. The relation is also patent when all debris discs with a solid gas detection are included: all eight stars in Table~\ref{DDgas} are early spectral type stars (B9 to A6).

The presence of gas does not seem to be influenced by the strength of the infrared excess: HD 181327 shows an infrared excess that is more than three times that of HD 172555, yet it is not detected; HD 1624249 also shows a prominent excess, larger than the infrared excess for both HD 172555 and HD 181296; but again it shows no gas emission within the sensitivity of Herschel-PACS. The continuum emission at 160 $\rm \mu m$ towards HD 172555 and HD 181296 is 0.031 and  0.111 Jy, respectively, compared to 0.850 Jy towards HD 181327 and 0.249 Jy towards HD 164249, so if continuum plays a major role in gas emission both HD 164249 and HD 181327 would have been easily detected. 

Considering the variety of gas properties in BPMG, it seems that different mechanisms can operate at the same age, and stochastic events may play an important role by producing huge amounts of gas in a short period of time while a more quiescent gas production results from grain dust collisions, evaporation and cometary activity. Future research is needed to understand the gas production mechanisms in debris discs, and the identification of observables useful for distinguishing the various mechanisms is a main goal for future work. The detection of other gaseous species towards \object{HD~172555} and \object{HD~181296} will help us to understand the origin of the gas. Such observations can be made with ALMA, with the aim of detecting species like SiO, CO, and C I.

\section{Summary and conclusions}
We observed 19 BPMG members with the \textit{Herschel Space Observatory} instrument PACS. Our main conclusions follow.

1. We have detected infrared emission in excess at 70/100/160 $\mu \rm{m}$ in 8/9/7 out of the 16/16/19 objects observed at each band. The detection ratio is greater than the typical value of 15--20 \% for older systems, but we highlight that our sample is small and not robust against selection effects.

2. We detected emission at 70 $\mu {\rm m}$ towards \object{AT~Mic}, \object{HD~146624}, and \object{HD~29391} for the first time \object{HD~29391}  was also detected at 100 $\rm \mu m$. The 70 $\mu {\rm m}$ excess flux in \object{HD~29391} results in a very low infrared fractional luminosity of $\rm L_{IR}/L_{*} \sim 2.0 \times 10^{-6}$. 

3. We modelled the dust emission with modified blackbody models. Blackbody temperatures range from 58 K (\object{HD~129391}) to 229 K (\object{HD~172555}). Dust masses range from $\rm 6.6 \times 10^{-5}~M_{\oplus}$ to  $\rm 0.20~M_{\oplus}$ and black body estimates for the inner radii range from 2 to $\rm \sim 82~AU$. 

4. The object HD 203 shows an unusually steep SED in the 70-100 $\rm \mu m$ range that can be a signpost of an under-abundance of large grains.

5. One of the systems studied, \object{HD~172555}, shows [OI] emission at 63 $\mu \rm{m}$. Another of the systems studied, HD 181296, shows [CII] emission at 157 $\rm \mu m$.

\object{HD~172555} and HD 181296 are both warm debris discs, and we note that the presence of gas could be linked to hot dust, in such a way that the same phenomenon producing the gas also produces a population of small grains near the star.

\acknowledgements
We acknowledge the anonymous referee for a very interesting discussion that helped improve the quality of the paper. This research has been funded by Spanish grants BES-2008-003863, AYA 2010-21161-C02-02, AYA 2012-38897-C02-01, AYA 2011-26202,   and PRICIT-S2009/ESP-1496. I. Kamp and P. Rivi\'{ere}-Marichalar acknowledge funding from an NWO MEERVOUD grant. We also acknowledge support from ANR (contract ANR-07-BLAN-0221) and PNPS of CNRS/INSU, France. C. Pinte acknowledges funding from the
European Commission's 7$^\mathrm{th}$ Framework Programme
(contract PERG06-GA-2009-256513) and from
Agence Nationale pour la Recherche (ANR) of France under contract
ANR-2010-JCJC-0504-01.

\bibliographystyle{aa} % style aa.bst 2
\bibliography{biblio.bib}
\end{document}